\documentclass[a4paper,12pt]{article}
\usepackage[warn]{mathtext}				
\usepackage[T1, T2A]{fontenc}			
\usepackage[utf8]{inputenc} 					 		
\usepackage[utf8]{inputenc} 			 
\usepackage{wrapfig}				
\usepackage{amssymb, amsmath}			 
\usepackage{indentfirst}   
\usepackage[titletoc]{appendix}
\usepackage{mathrsfs}
\usepackage{cmap} 			
\usepackage[pdftex]{graphicx}
\usepackage{pgfplotstable}	
\usepackage[bottom]{footmisc}
\usepackage[hidelinks]{hyperref}
\DeclareGraphicsExtensions{.eps}
\newcommand*{\defeq}{\stackrel{\text{def}}{=}}
\usepackage{graphicx}

\usepackage{float}
\usepackage[left=20mm,right=20mm,top=20mm,bottom=20mm,bindingoffset=0mm]{geometry}
\let\DS     = \displaystyle

\iffalse

\else

\title{Energy supply into a semi-infinite $\beta$-Fermi-Pasta-Ulam-Tsingou chain by periodic force loading}

\author{Sergei D. Liazhkov$^{1,2}$}
\date{%
    $^1$Peter the Great Saint-Petersburg Polytechnic University, St. Petersburg, Russia\\%
    $^2$Institute for Problems in Mechanical Engineering of the Russian Academy of Sciences, St. Petersburg, Russia.\\[4mm]%
    \today
}
\def\edited#1{{\textcolor{black}{#1}}}

\begin{document}

\maketitle

\begin{abstract}
  We deal with dynamics of the~$\beta$-Fermi-Pasta-Ulam-Tsingou chain with one free end, subjected to the~\edited{sudden} sinusoidal force. We examine the evolution of the total energy supplied into the chain at large times. In the harmonic case~($\beta=0$), the energy grows in time linearly \edited{at the driving frequencies, corresponding to nonzero group velocities. The rate of energy supply is shown to decrease with increasing excitation frequency. Loading with the cut-off frequency, corresponding to zero group velocity, results in the energy growth proportionally to~$\sqrt{t}$}. Explanation of behavior in time of the energy is proposed by analysis of the obtained approximate closed-form expression\edited{s} for the \edited{particle velocity}.  In the weak anharmonic case, large-time asymptotic approximation for the total energy is obtained by using the renormalized dispersion relation. \edited{The approximation allows one to estimate rate of the energy supply at the driving frequencies, which belong both in the pass-band and in the stop-band of the harmonic chain.}  Consistency of the asymptotic \edited{estimates} with the results of numerical simulations is discussed.
\end{abstract}

\section{Introduction}\label{sect1}

Investigation of \edited{the} response of a medium under external excitation is one among the oldest engineering problems, with which one has been encountered~\cite{Pupin, mead1971}. Nowadays the problem concerns not merely the macroscale, but rather the micro- and nanoscale, wherein one deals with discrete structures (see, e.g.,~\cite{DNA, Sinko}).
\par
Wave transmission \edited{to} the discrete linear medium is known to be possible only at driving frequencies, enclosed \edited{within} the pass-band\edited{s}, \edited{determined by the dispersion relation}. In contrast, energy can propagate into the nonlinear medium at frequencies outside the pass band\edited{s, i.e., in the stop-bands}. It was shown, e.g., in~\cite{Ovch, Sievers, Flach, Kosevich, Dolgov, Sato} that nonlinear interactions result in the nucleation of the intrinsic localized modes~(ILMs)\footnote{Often in literature ILMs are referred either to as the discrete breathers, see, e.g.,~\cite{Flach} or to as the self-localized modes, see, e.g.,~\cite{Kosevich}.}. It was shown in~\cite{PREBreathers, eva2017, Caputo} that energy propagation, \edited{induced} by ILMs, occurs at their frequencies, \edited{which may lie both below and above the pass-band~\cite{PREBreathers} and also in the gap between the pass-bands~\cite{eva2017,Caputo}}. \edited{The} experiments of Watanabe~\cite{Watanabe2012, Watanabe2015, Watanabe2017, Watanabe2018} showed the \edited{phenomenon} of supratransmission, namely a sudden energy flow into the discrete media at the frequency in the stop-band due to ILMs, provided that some threshold value of the excitation amplitude is reached. \edited{Extensive studying of the supratransmission is possibly explained by its proposed applications, e.g., to the Josephson contacts~\cite{Leon, Santis}, optical waveguides~\cite{Khomeriki}, digital amplifiers of the weak signals~\cite{Khomeriki2} etc.}
\par
There are plenty of works, devoted to theoretical analysis of the energy propagation, induced by the supratransmission. In general, theoretical studies are focused on determining of conditions of the energy propagation by ILMs, with which emergence of the supratransmission is associated. The bibliography of such studies starts, probably, from the paper~\cite{Geniet2002}, wherein the supratransmission was shown to be accompanied by sharp increasing of the supplied energy. This process was described by dint of the~\edited{sine}-Gordon equation, which became hereinafter a model equation for studying of the supratransmission~(in particular, in the Josephson junctions) both in the discrete~(see, e.g., \cite{Geniet2003, Leon, Macias2008}) and in the continuum~(see, e.g., \cite{Macias2016, Santis, Santis1, Santis2}) formulations. Similar results  were obtained for the models obeying the \edited{discrete} nonlinear Schr{\"o}dinger equation~\cite{Susanto, Motcheyo2}, for FPUT chains~\cite{Khomeriki,Macias2018, Macias2020}, electrical lattices~\cite{Kenmogne, Motcheyo} and others~\cite{Gendelman2022}. On the other hand, evolution of the energy, transmitted into the lattice by the loading at driving frequencies, lying both in the stop-band and also in the pass-band, is the rarely studied problem. In the paper~\cite{PREBreathers}, evolution of the energy, transmitted into the infinite chain with on-site potential interactions and supplied by kinematic loading, was demonstrated. The non-stationary processes of the energy supply may be described analytically in the harmonic approximation, what was done for the infinite harmonic chain on the linear elastic substrate~\cite{Kuz2018} for the cases of both kinematic and force loadings. For small driving amplitudes, \edited{the} rate of supply of the total energy into the nonlinear chain is shown~\cite{PREBreathers} to be indistinguishable from the one described in the harmonic approximation.
\par In the present paper, we deal with \edited{the} influence of the free boundary on the rate of energy supply. Some results, \edited{regarding} the latter, were obtained in~\cite{Cannas1991} for the case of kinematic loading~(numerically) and in~\cite{mokole1990exact} for the case of force loading. To the best of our knowledge, there has not been any description of process of the energy supply into anharmonic lattices yet. We address this issue in the current work, considering the \edited{semi-infinite free end}~$\beta$-FPUT monoatomic chain\footnote{We note that the potential itself is unphysical, although the~$\beta$-FPUT models are systematically used for \edited{estimation} of manifestation of nonlinearity on the dynamical processes and, for this reason, we consider interactions~(between the particles) via this potential here. We suppose that obtained in the current paper results, concerning the process of energy supply, will improve comprehension of the latter for a possible generalization of the results to cases of the interactions via more realistic potentials.}. We manage here to obtain analytically some estimates of the weak anharmonicty on the processes of energy supply.  The model presented in the current work is suitable for the experimental setup, described in~\cite{Watanabe2017, Watanabe2018} and is thus expected to be validated by means of \edited{the aforementioned} experiments and \edited{to be auxiliary to provide a description of more complicated experiments~\cite{cherednichenko2019nonlinear}.} However, our main motivation for the current study is closely related to problems, concerning nanoscale heat transport, to which much of theoretical studies~(see, e.g.,~\cite{Dhar, ChenG, Lepri2016, Podolskaya, Lepri2023, ChenJ}) are devoted. \edited{Indeed, the results of the present paper are supposed to be auxiliary for problems, which have application to devices for control heat fluxes in nanomaterials, namely on heat transport in inhomogeneous lattices~(see Sect.\ref{Concl} for details) and also in the force driven lattices~\cite{ai2010heat}.}

The paper is organized as follows. \edited{In Sect.~\ref{sect1_5}, we discuss the notation.} In Sect.~\ref{sect2}, we formulate the problem. In Sects.~\ref{VEL}--\ref{H}, we consider the problem in the harmonic approximation, namely, for the Hooke chain\footnote{This is the monoatomic harmonic chain of identical particles, connected by the linear identical springs, see~\cite{ZAMM}.}. In particular, in Sect.~\ref{VEL1}, we obtain the equation for the field of particle velocities analytically, by using which we perform \edited{an} asymptotic estimation for the total energy in Sect.~\ref{H}. For explanation of \edited{the} behavior in time of the total energy and \edited{the} peculiarities of transmission of the latter into the chain, we find an approximate closed-form solution for the particle velocity~(Sect.~\ref{VEL2}). We underline that \edited{the} main results of Sects.~\ref{VEL}--\ref{H} are not new, but rather~\edited{re-obtained and, besides, are compared with the corresponding numerical calculations}. The \edited{analytical} results, obtained in Sect.~\ref{VEL2}, were previously got in~\cite{mokole1990exact}, but the final results of this section are refined. The total energy, supplied into the chain, was previously estimated in~\cite{mokole1990exact} but \edited{loading} by the cut-off frequency was not analyzed and we consider \edited{this case} in Sect.~\ref{H}. The main novelty of the present paper is in Sect.~\ref{AH}, where we investigate \edited{the} influence of anharmonicity on \edited{the} behavior of the supplied energy. The asymptotic approximation for the total energy, obtained in Sect.~\ref{AH1}, is analyzed for the driving frequencies, lying both inside the pass-band~(Sect. \ref{AH2}) and inside the stop-band~(Sect.~\ref{AH3}) of the Hooke chain. In Sect.~\ref{Concl}, results of the paper and their possible generalizations are discussed.

\edited{\section{Nomenclature}\label{sect1_5}
In the paper, we use the following notation:}\\[2mm]
\par $\mathbb{N}$ is the set of all natural numbers;
\par $t$ is the time;
\par $n$ is a particle number;
\par $\dot{(...)}$ and~$\ddot{(...)}$ stand for the first and second time derivatives respectively;
\par $H(...)$ is the Heaviside function;
\par $\delta_{n,j}$ is the Kronecker delta~($1$ if~$n=j$ and~$0$ otherwise);
\par $\theta$ is the wave number;
\par $\omega_e$ is the elementary atomic frequency;
\par $\epsilon$ is a small quantity with dimension of $\mathrm{s}^{-1}$;
\par $\mathcal{T}_n$ is the Chebyshev polynomial of the first kind;
\par $F_0$ is the amplitude of the driving force;
\par $\Omega$ is the driving frequency.

\section{Problem statement}\label{sect2}

We consider forced oscillations of the semi-infinite chain, consisting of identical particles with mass~$m$, which interact with the nearest neighbors via the hard-type~$\beta$-Fermi-Pasta-Ulam-Tsingou potential. The chain has one free end, excited by the sudden periodic force with the amplitude~$F_0$ and the frequency~$\Omega$. The governing dynamical equations are 
\begin{equation}\label{DynEQ}
    \begin{array}{l}
     \dot{u}_n=v_n,\\[2mm]
         m\dot{v}_0=c(u_{1}-u_{0})+\beta(u_{1}-u_{0})^3+F_0\sin (\Omega t)H(t),\\[2mm]
     m\dot{v}_n=c(u_{n+1}-2u_n+u_{n-1})+\beta \left((u_{n+1}-u_n)^3-(u_n-u_{n-1})^3\right), \quad n\in \mathbb{N},
    \end{array}
\end{equation}
where~$H(t)$ is the Heaviside function;~$u_n$ and~$v_n$ are displacement and velocity for particle~$n$ respectively; $c$ and~$\beta$ are harmonic and anharmonic force constants respectively. We assume that interatomic interactions are associated with weak anharmonicity, i.e.,
\begin{equation}\label{Cond_beta}
    \frac{\beta F_0^2}{c^3}\ll 1.
\end{equation}
Initially, the chain is unperturbed, i.e.,
\begin{equation}\label{IC}
   u_n=0,\qquad \dot{u}_n=0.
\end{equation}
In the next sections, we investigate rate of energy supply into the chain as function of the driving frequency~$\Omega$ and range of the frequencies, permitting the energy transmission. In order to have a benchmark for the anharmonic case, we address the issue on energy supply in the harmonic approximation~(for~$\beta=0$) below. 

\section{Field of particle velocities in the harmonic chain}\label{VEL}
\subsection{Exact solution}\label{VEL1}
Reformulating the problem statement for the Hooke chain, rewrite the equations of motion~(\ref{DynEQ}) as follows:
\begin{equation}\label{EQ_harm}
\begin{array}{l}
    \DS \dot{v}_n=\omega_e^2(u_{n+1}-u_n)-\omega_e^2(u_n-u_{n-1})(1-\delta_{n,0})-2\eta v_n+\frac{F_0}{m}\sin (\Omega t)H(t)\delta_{n,0},\\[2mm]
    \omega_e=\sqrt{c/m},\quad \eta \ll \omega_e,
\end{array}
\end{equation}
where~$\omega_e$ is the elementary atomic frequency; $\delta_{n,j}$ is the Kronecker delta; $\eta$ is the kinematic viscosity of the external medium. Here, we intend to add the term~$-2\eta v_n$ into the equations of motion and, following the limiting absorption principle~(see, e.g.,~\cite{Indeitsev1999}), limit the obtained solution to the one for initially considered non-dissipative case~($\eta\rightarrow0+$).
Initial conditions are rewritten in the generalized form~\cite{Vladimirov}:

\begin{equation}\label{IC2}
    u_n\vert_{t<0}=0.
\end{equation}
Equations~(\ref{EQ_harm}) describe forced oscillations of the semi-infinite chain. The problem of the latter was solved analytically in~\cite{Cannas1991, mokole1990exact, Lee, Lee1972, Nay, Prato}. Kinematic loading was considered in the form of sinusoidal law in~\cite{Cannas1991, Prato} and in the form of the linear law in~\cite{Nay}~(loading with constant velocity). In~\cite{mokole1990exact, Lee, Lee1972} force loading was considered. However, the methods, which were applied in the aforesaid studies, seem cumbersome.

In order to solve Eqs.~(\ref{EQ_harm}), we introduce the direct and inverse discrete cosine transforms (DCT)~\cite{Ahmed}:
\begin{equation}\label{DCT}
    \DS \hat{u}(\theta)=\sum_{n=0}^\infty u_n\cos \frac{(2n+1)\theta}{2},\quad u_n=\frac{1}{\pi}\int_{-\pi}^\pi \hat{u}(\theta)\cos \frac{(2n+1)\theta}{2}\mathrm{d}\theta,\quad \hat{u}\vert_{t<0}=0,
\end{equation}
where~$\theta$ is the real-valued wavenumber; $\hat{u}$ is some time-dependent function.\\[1mm]
Applying~DCT~(\ref{DCT}) to Eq.~(\ref{EQ_harm}) yields 
\begin{equation}\label{IM_ODE}
\begin{array}{l}
   \ddot{\hat{u}}+2\eta\dot{\hat{u}}+\omega^2 \hat{u}=\DS \frac{F_0}{m}\sin (\Omega t)H(t)\cos\frac{\theta}{2},\quad \omega(\theta)=\DS 2\omega_e \Big\vert\sin \frac{\theta}{2}\Big\vert, \\[2mm]
\end{array}
\end{equation}
where~$\omega(\theta)$ is the dispersion relation for the Hooke chain, corresponding to the pass-band, whence\footnote{Eq. for~$\hat{u}$ is calculated by the Duhamel integral: convolution of the fundamental solution of~\edited{the} homogeneous part of Eq.~(\ref{IM_ODE}) and its right part.}
\begin{equation}\label{IMAG1}
\hat{u}=H(t)\left(\hat{u}_{\omega}+\hat{u}_{\Omega}\right)\cos\frac{\theta}{2},
\end{equation}
\begin{equation}\label{IMAG2}
    \DS\hat{u}_{\omega}=\frac{F_0\Omega \sqrt{4\eta^2(\omega^2-\eta^2)+(\Omega^2-\omega^2+2\eta^2)^2}}{m((\Omega^2-\omega^2)^2+4\eta^2\Omega^2)\sqrt{\omega^2-\eta^2}}e^{-\eta t}\sin \left(t\sqrt{\omega^2-\eta^2}+\phi_1 \right),
\end{equation}
\begin{equation}\label{IMAG3}
    \DS \hat{u}_{\Omega}=\frac{F_0\sin(\Omega t-\phi_2)}{m\sqrt{(\Omega^2-\omega^2)^2+4\eta^2\Omega^2}},
\end{equation}
\begin{equation}\label{IMAG4}
    \phi_1(\theta)=\arccos \left(\frac{\Omega^2-\omega(\theta)^2+2\eta^2}{\sqrt{4\eta^2(\omega(\theta)^2-\eta^2)+(\Omega^2-\omega(\theta)^2+2\eta^2)^2}}\right),
\end{equation}
\begin{equation}\label{IMAG5}
    \phi_2(\theta)=\arccos \left(\frac{\omega(\theta)^2-\Omega^2}{\sqrt{(\Omega^2-\omega(\theta)^2)^2+4\eta^2\Omega^2}}\right).
\end{equation}
Applying inverse DCT to~$\dot{\hat{u}}$ yields expression for the particle velocity:
\begin{equation}\label{Vel1}
\DS v_n=v_n^\omega+v_n^\Omega,
\end{equation}
\begin{equation}\label{Vel2}
    \DS v_n^\omega=\frac{H(t)}{\pi}\int_{-\pi}^{\pi}\dot{\hat{u}}_\omega(\theta)\cos\frac{(2n+1)\theta}{2} \cos \frac{\theta}{2}\mathrm{d}\theta,
\end{equation}
\begin{equation}\label{Vel3}
    \DS v_n^\Omega=\frac{H(t)}{\pi}\int_{-\pi}^{\pi}\dot{\hat{u}}_{\Omega}(\theta) \cos\frac{(2n+1)\theta}{2} \cos \frac{\theta}{2}\mathrm{d}\theta,
\end{equation}
written in the integral form. The Eq.(\ref{Vel1}) is represented as exact solution of the dynamical problem~(\ref{EQ_harm}) and is divided into two terms, corresponding to the the stationary~(forced) and decaying natural oscillations~($v_n^\Omega$ and~$v_n^\omega$ respectively) and is further employed to obtain the total energy, supplied into the chain~(see Sect.\ref{H1}). However, the integral representation for the particle velocity does not allow to understand nature of propagating into the chain waves clearly. Therefore, we derive an approximate solution for the particle velocity in a closed form below.
\subsection{Approximate solution}\label{VEL2}
 Here, we obtain an approximate solution of the dynamical equations~(Eq.~(\ref{EQ_harm})) in the limit~$\eta \rightarrow 0+$, considering every term in~(Eq.~(\ref{Vel1})), corresponding to the forced and natural oscillations, separately. \edited{Probably} for the first time this approach was implemented by Hemmer~(for the particle displacement in the conservative~\edited{infinite} Hooke chain, see~\cite{Hemmer}) by obtainment of the stationary solution~(at~$t\rightarrow \infty$). An approach, which is \edited{introduced} below, aims to \edited{refine the approximate solution by analysis of the non-stationary solution of the dynamical problem}. 
 
 Eq. for~$v_n^\Omega$ is obtained in the following closed form~(see Appendix~\ref{AAA}):
 \begin{equation}\label{ForcedContr}
v^\Omega_{n}=\left[\begin{gathered}\DS\frac{(-1)^{n+1}F_0}{m\omega_e}\mathcal{T}_{2n+1}\left(\DS\frac{\Omega}{2\omega_e}\right)\cos(\Omega t)H(t), \quad 0<\Omega\leqslant 2\omega_e,\\[3mm]
\DS\frac{(-1)^{n+1}F_0(1-e^{-\gamma})}{m\sqrt{\Omega^2-4\omega_e^2}}e^{-\gamma n}\cos(\Omega t)H(t), \quad  \Omega>2\omega_e,\;\DS\gamma=2\mathrm{arccosh}\frac{\Omega}{2\omega_e},\end{gathered}\right.
\end{equation}
where~$\mathcal{T}_{n}(x)\defeq\cos(n\arccos{x})$ is the Chebyshev polynomial of the first kind. For~$\Omega>2\omega_e$, Eq.(\ref{ForcedContr}) can also be derived by differentiation with respect to the time of Eq. (34) in~\cite{mokole1990exact} or Eq.(8) in~\cite{Cannas1991} taking into account the case of force loading. The latter procedure leads to the same results, written in another form. The quantity~$\gamma$ is the attenuation coefficient~(or inverse of localization length, see~\cite{Cannas1991}), which coincides with the imaginary part of the complex wave number, corresponding to the dispersion relation for the stop-band of the Hooke chain~(for $\omega>2\omega_e$, see~\cite{Gavr2023, shishkina2023localized}) and determined at the frequency~$\Omega$.

The expression for~$v_n^\omega$ is asymptotically estimated at large times~($\Omega t\gg 1$) as
\begin{equation}\label{NatContr}
    v_n^\omega\backsim v^\omega_{n,\mathrm{cr}}+ v^\omega_{n,\mathrm{st}},
\end{equation}
where~$\backsim$ stands for the sign of the large-time asymptotically equivalent function. Here~$v^\omega_{n,\mathrm{cr}}$ is large time asymptotics of~$v_n^\omega$ in a form of the contribution from the critical\footnote{This point becomes singular at~$\eta\rightarrow 0+$.} point of the integrand in Eq.(\ref{Vel2})~(see Appendix~\ref{BBB}):
\begin{equation}\label{CritEq}
v^\omega_{n,\mathrm{cr}}=\left[\begin{gathered}
     \DS\frac{F_0\mathcal{T}_{2n+1}\left(\sqrt{1-\frac{\Omega^2}{4\omega_e^2}}\right)}{m\omega_e}\sin(\Omega t), \quad 0<\Omega< 2\omega_e,\\[2.5mm]
     \DS\frac{(-1)^nF_0(2n+1)}{m\sqrt{\pi\omega_e^3t}}\sin \left(2\omega_e t+\frac{\pi}{4}\right), \quad \Omega=2\omega_e.
    \end{gathered}\right.
\end{equation}
For~$\Omega>2\omega_e$, $v^\omega_{n,\mathrm{cr}}\equiv 0$. Note that, for~$0<\Omega\leqslant 2\omega_e$, the contributions~(\ref{ForcedContr}) and~(\ref{CritEq}) themselves are bereft of physical meaning
in contrast to their sum, written in the generalized sense~(see~expression~(\ref{VEL23_5})). The contribution~$v^\omega_{n,\mathrm{st}}$ is the large time asymptotics of~$v^\omega_n$ at the moving front~(see Appendix~\ref{CCC}):
\begin{equation}\label{StatEq}
v^\omega_{n,\mathrm{st}}=\frac{2F_0\Omega H(\Tilde{n}_s-n)nt}{\sqrt{\pi}m\omega_e\sqrt[4]{\omega_e^2t^2-n^2}\left(\left(\Omega^2-4\omega_e^2\right)t^2+4n^2\right)}\sin \left(\psi_nt+\hat{\varphi}_n-\frac{\pi}{4}\right),
\end{equation}
\begin{equation}\label{StatEq2}
\psi_n(t)\defeq\frac{2\sqrt{\omega_e^2t^2-n^2}}{t}-\frac{2n}{t}\arccos{\frac{n}{\omega_et}},\quad \hat{\varphi}_n(t)\defeq \arcsin {\frac{n}{\omega_et}}\quad \Tilde{n}_s=\frac{v_st}{a},
\end{equation}
where~$a$ is the undeformed bond length; ~$v_s \defeq \omega_ea$ is the speed of sound. Here~$\Tilde{n}_s$ is referred to as a coordinate of a point moving with the speed of sound~\edited{and~$\psi_n(t)$, defined in Eq.(\ref{StatEq2}) is referred to as an effective frequency of perturbation, propagating with the same speed.}
\par Using the contributions~(\ref{ForcedContr}), (\ref{CritEq}), (\ref{StatEq}), we construct an approximate solution for the particle velocity. Here~(and further), in order to check accuracy of the obtained below approximations, we integrate the dynamical equations~(\ref{EQ_harm}) in the dimensionless form~(putting~$\eta=0$) with zero initial conditions using the symplectic leap-frog method with the time step~$0.01/\omega_e$.  The chain with~$5\cdot 10^3$ particles is considered with the force boundary condition for the left end and with the fixed boundary condition for the right end\footnote{A boundary condition for the right end plays no role for the large enough number of particles.}. Below we consider three cases regarding to the driving frequency~$\Omega$.

\subsubsection{Case $\Omega>2\omega_e$}\label{APP_MORE}
\begin{figure}[htb]
\center{\includegraphics[width=0.75\linewidth]{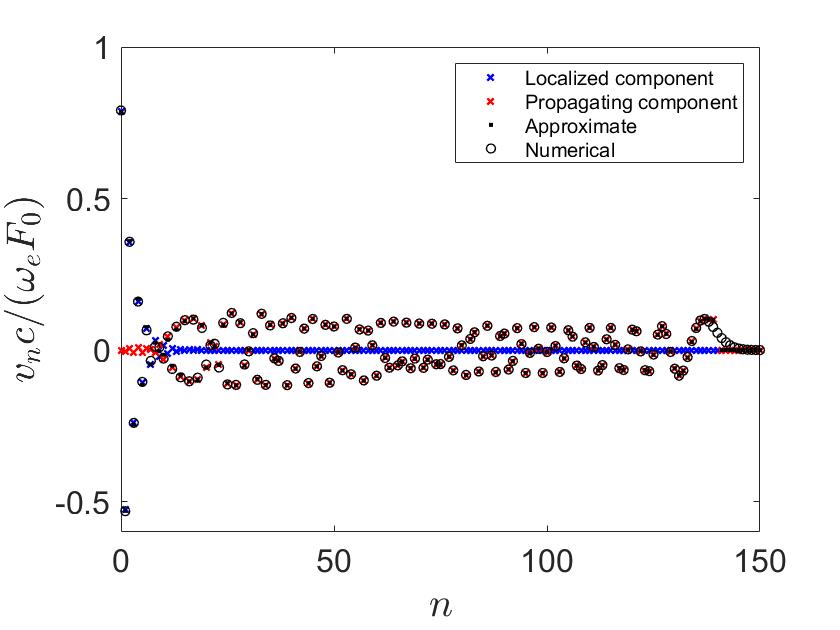}}
\caption{Comparison of the \edited{analytical}  solution~(expression~(\ref{VEL23_6})) for the particle velocity with the corresponding numerical solution. \edited{The results are shown} at~$\omega_et=140$ and for~$\Omega=2.04\omega_e$.  The crosses and stars correspond to the contributions~$ v^\omega_{n,\mathrm{st}}$ and~$v_n^\Omega$ respectively. }
\label{vel_out}
\end{figure}

In this case, the solution for particle velocity has the simple form
\begin{equation}\label{VEL23_6}
    v_n\approx  v^\omega_{n,\mathrm{st}}+v_n^\Omega,
\end{equation}
where~$v^\omega_{n,\mathrm{st}}$ and~$v_n^\Omega$ are determined by the expressions~(\ref{StatEq}) and~(\ref{ForcedContr}) respectively. The first term, $v^\omega_{n,\mathrm{st}}$,  corresponds to the contribution from the propagating waves, which decays in time as~$1/t^{3/2}$~(for the points far from the boundary but not close to the wavefront) and the second term, $v_n^\Omega$, corresponds to the stationary localized oscillations near the driven boundary, which vanish with increasing of distance from the boundary~\edited{(zeroth particle)}. Thus, at the considered driving frequencies, the energy is impossible to be transmitted into the chain upon reaching some value. Approximation of the field~$v_n$ at the fixed moment of time is shown in Fig.~\ref{vel_out}.
It is seen in Fig.~\ref{vel_out} that the approximate solution practically coincides with the results of numerical solution, except at the points near the wavefront. \par

\subsubsection{Case $0<\Omega<2\omega_e$}\label{APP_LESS}
\begin{figure}[htb]
\begin{minipage}{0.5\linewidth}
\center{\includegraphics[width=0.95\linewidth]{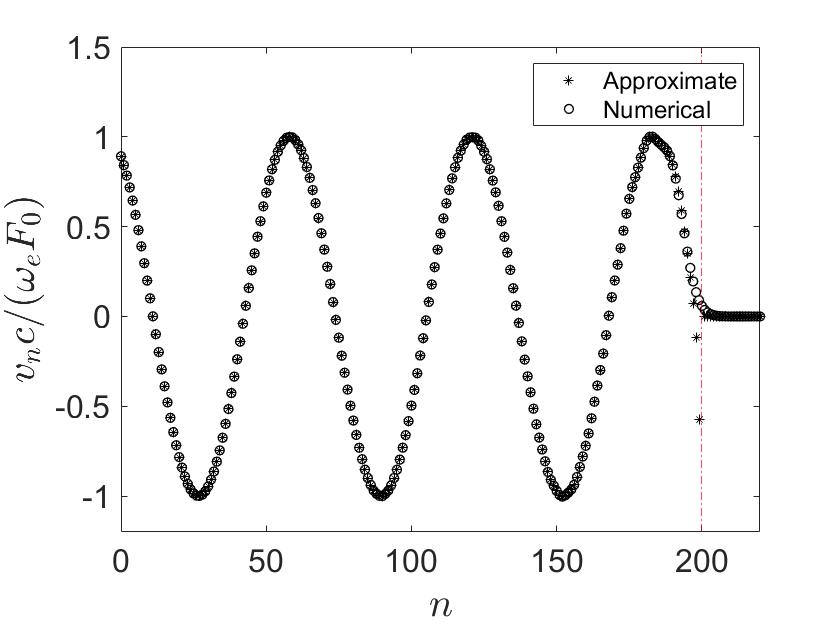}\\A }
\end{minipage}
\begin{minipage}{0.5\linewidth}
\center{\includegraphics[width=0.95\linewidth]{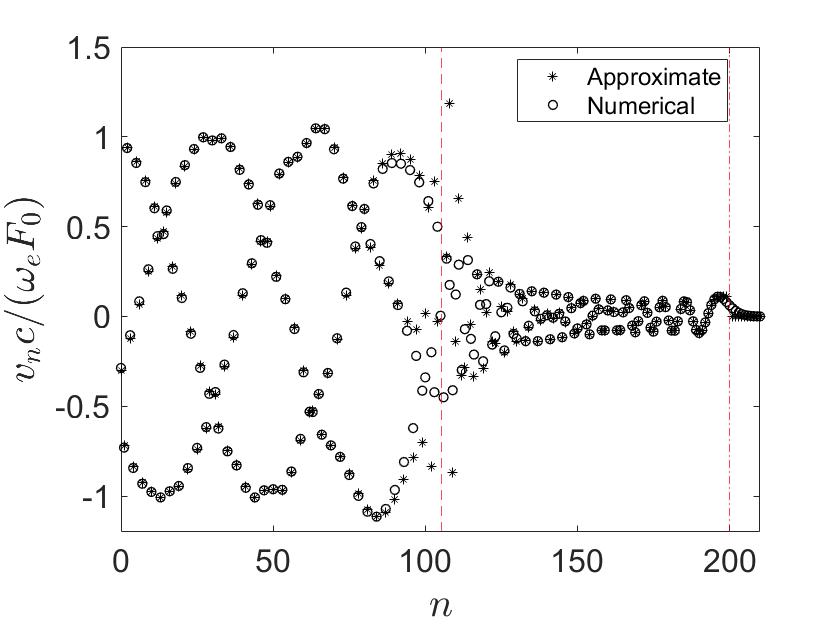} \\B}
\end{minipage}
\caption{Comparison of the \edited{analytical} solution~(expression~(\ref{VEL23_5})) for the particle \edited{velocity} with the corresponding numerical solution. \edited{The results are shown at}~$\omega_et=200$ and for the driving frequencies~$\Omega=0.1\omega_e$~(A) and~$\Omega=1.7\omega_e$~(B). The dashed and dash-dotted lines correspond to the coordinates of~$\Tilde{n}_g$ and~$\Tilde{n}_s$ respectively. }
\label{vel_inside_1}
\end{figure}

Before constructing the solution for this case, we note that the expression for the contribution~$v^\omega_{n,\mathrm{st}}$ for~~$0\leqslant \Omega\leqslant 2\omega_e$ has a singularity. Introduce a real-valued coordinate~$\Tilde{n}_g$, at which the contribution is singular. From~(\ref{StatEq}), we express the coordinate as 
\begin{equation}\label{VEL23_1}
    \Tilde{n}_g=\omega_et\sqrt{1-\frac{\Omega^2}{4\omega_e^2}}.
\end{equation}
Note that the group velocity for the Hooke chain is determined as
\begin{equation}\label{VEL23_2}
    v_g(\theta)\defeq a\frac{\mathrm{d}\omega}{\mathrm{d}\theta}=v_s\cos\frac{\theta}{2}\mathrm{sgn}\left(\sin\frac{\theta}{2}\right),
\end{equation}
or, as the function of the dispersion relation:
\begin{equation}\label{VEL23_3}
    v^g(\omega)=v_s\sqrt{1-\frac{\omega^2}{4\omega_e^2}}\mathrm{sgn}\,\omega.
\end{equation}
Therefore, we can rewrite Eq.~(\ref{VEL23_1}) as
\begin{equation}\label{VEL23_4}
    \Tilde{n}_g=\frac{v^g(\Omega)t}{a}.
\end{equation}
For~$0<\Omega<2\omega_e$, $\Tilde{n}_g>0$ is the coordinate of a point, moving with the group velocity, corresponding to the frequency~$\Omega$. We consider \edited{the} existence of a solution for the contributions to the particle velocity both behind the point~$\Tilde{n}_g$ and between~$\Tilde{n}_g$ and~$\Tilde{n}_s$. In the domain~$\Tilde{n}_g<n<\Tilde{n}_s$, the solutions for~$v^\omega_{n,\mathrm{cr}}$ and~$v^\Omega_n$ cannot exist, because~$\psi_{\Tilde{n}_g<n<\Tilde{n}_s}<\Omega$ for all time of motion of~$\Tilde{n}_g$ and thus wave with the frequency~$\Omega$ cannot propagate. Based on aforesaid, we construct the following approximate solution for~$v_n$:
\begin{equation}\label{VEL23_5}
    v_n\approx v^\omega_{n,\mathrm{st}}+\left(v_n^\Omega+v^\omega_{n,\mathrm{cr}}\right)H(\Tilde{n}_g-n),
\end{equation}
where~$v^\omega_{n,\mathrm{cr}}$ is determined by the expression~(\ref{CritEq}). Thus, there is an additional wavefront, which propagates with the group velocity, corresponding to the frequency~$\Omega$, besides one, propagating with the speed of sound. In contrast to the first term, which decays, the second term performs non-decaying oscillations with the frequency~$\Omega$. Note that expression for the second term in~(\ref{VEL23_5}),~$v_n^\Omega+v^\omega_{n,\mathrm{cr}}$, is \edited{the} known result: the latter can be derived by differentiation with respect to time of Eq.~(34) in~\cite{mokole1990exact}. In that work, the latter is understood as a solution of dynamical problem at large times. Here, we refine the latter by addition of the contribution~$v^\omega_{n,\mathrm{st}}$ and by constraints on the existence of the term~$v_n^\Omega+v^\omega_{n,\mathrm{cr}}$, which we calculate by using another method, based on the limiting absorption principle. Approximation of the field~$v_n$ at the fixed moment of time is shown in Fig.~\ref{vel_inside_1}. 

\par
It is seen from Fig.~\ref{vel_inside_1}A that the velocity field is practically sinusoidal. Since the driving frequency is rather small~($\Omega \ll \omega_e$), the front~$\Tilde{n}_g$, carrying the energy, practically coincides with the front~$\Tilde{n}_s$, moving with the speed of sound. The effect of the dispersion can be negligible. For high driving frequencies, the case is quite another~(see Fig.~\ref{vel_inside_1}B). The front~$\Tilde{n}_g$ moves significantly slower than~$\Tilde{n}_s$, what is shown below to affect \edited{on} the energy transfer into the chain. It is seen from Fig.~\ref{vel_inside_1} that the approximation for the velocity field, obtained here, becomes inadequate in the neighborhood of the wavefronts. The accurate estimation of the velocity field for~$n=\lfloor\Tilde{n}_g\rfloor+O(1)$ and~$n=\Tilde{n}_s$ is beyond the scope of present paper. If necessary, the latter can be performed by operation of the proposed in~\cite{Gavr2023, Gavr1999} methods. \par

\subsubsection{Case $\Omega=2\omega_e$}\label{APP_JUST}
\begin{figure}[htb]
\begin{minipage}{0.5\linewidth}
\center{\includegraphics[width=0.95\linewidth]{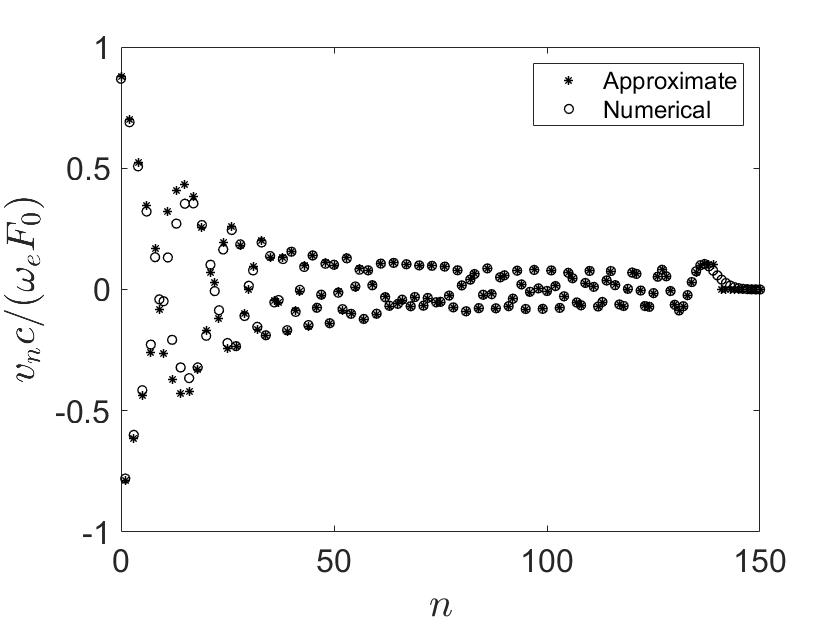} \\A }
\end{minipage}
\begin{minipage}{0.5\linewidth}
\center{\includegraphics[width=0.95\linewidth]{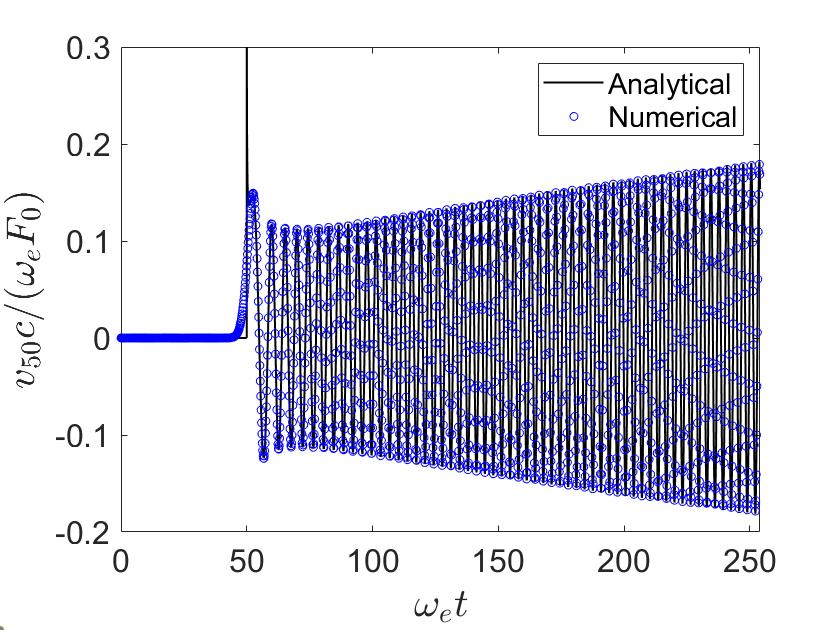} \\B}
\end{minipage}
\caption{A: Comparison of the \edited{analytical} solution~(Eq.(\ref{VEL23_7}) at~$h=10$) for the particle velocity with the corresponding numerical solution. \edited{The results are shown at~$\omega_et=140$ and for~$\Omega=2\omega_e$}. B: Evolution in time of the~$50$-th particle velocity at~$\Omega=2\omega_e$. \edited{Comparison of the analytical~(Eq.~(\ref{VEL23_8})) and numerical solutions is shown.}}
\label{vel_cutoff}
\end{figure}

In this case, ~$\Tilde{n}_g=0$ and, therefore, the contribution~$v^\omega_{n,\mathrm{st}}$ is singular~(and thus inaccurate) at the boundary. However, it is seen that the second expression in~(\ref{CritEq}) becomes inaccurate for large enough~$n$. Therefore, an approximate solution can be constructed by a crosslinking of the two solutions at a point~$h$, which is not far from the free boundary:
\begin{equation}\label{VEL23_7}
    v_n\approx v^\omega_{n,\mathrm{st}}H(n-h)+\left(v_n^\Omega+v^\omega_{n,\mathrm{cr}}\right)H(h-n).
\end{equation}
Therefore, for~$1\ll n <\Tilde{n}_s$, the expression for the particle velocity,~$v_n$, can be written as
\begin{equation}\label{VEL23_8}
    v_n\approx v^\omega_{n,\mathrm{st}}= \frac{F_0H(\Tilde{n}_s-n)t}{\sqrt{\pi}m\sqrt[4]{\omega_e^2t^2-n^2}n}\sin\left(\psi_nt+\hat{\varphi}_n-\frac{\pi}{4}\right).
\end{equation}
 From the expression~(\ref{VEL23_8}) it is seen that, at~$\Omega=2\omega_e$, expression \edited{for the far field of particle velocity} grows in time as~$\sqrt{t}$. Note that approximate solution for the particle velocity in the infinite Hooke chain~(see time derivative of Eq.~(1.58) in~\cite{Hemmer}) also grows in time as~$\sqrt{t}$, however, with much larger amplitude than one in the semi-infinite Hooke chain, because the latter decreases with increasing~$n$. Comparison of approximate and numerical solutions is shown in Fig.~\ref{vel_cutoff}. 

It is seen in Fig.~\ref{vel_cutoff}\edited{A} that \edited{the analytical} and numerical solutions are in good agreement except the neighborhood of the point~$h$, in where the contributions, defined in the first and second terms in~Eq.~(\ref{VEL23_7}) are crosslinked.  Comparing of the approximations~(\ref{VEL23_7}) and~(\ref{VEL23_8}) with \edited{the} time derivative of Eq.(25) in~\cite{mokole1990exact}, we conclude that the latter result is erroneous. \edited{Dependence of the 50-th particle velocity is shown in Fig.~\ref{vel_cutoff}B. The approximate and numerical solutions are shown to coincide at large times. However, according to additional analysis of Eq.~(\ref{VEL23_8}), at extra large times the amplitude of the velocity gradually deviates from the one obtained numerically, although the approximate solution itself
behaves asymptotically closely to the numerical one.} 

\par Thus, at zero group velocity, the expression for the field of particle velocities grows in time\footnote{The similar effect is regularly observed in the continuum systems, see, e.g.,~\cite{SlepyanTsareva}.}. As \edited{shown in the following}, the latter is reason of growth of the total energy, although the contributions from the oscillations with the driving frequency and from the singular point are also localized near the boundary~(as at~$\Omega>2\omega_e$).

In the next section, we estimate the total energy, supplied into the chain.

\section{Energy transfer into the semi-infinite Hooke chain}\label{H}
\subsection{Exact equation for the energy}\label{H1}

According to the energy balance law, the total energy of the chain, $U$, is equal to the work done by the external force to displace zeroth particle:
\begin{equation}\label{WORK}
    U=F_0\int_0^t v_0(\tau)\sin(\Omega \tau)\mathrm{d}\tau.
\end{equation}
Substitution of~$\eta=0$ and~$n=0$ to~(\ref{Vel1}) and then~(\ref{Vel1}) to~(\ref{WORK}) using evenness property of the integrands yields
\begin{equation}\label{WORK_EXACT}
\begin{array}{l}
   \DS U= \DS \frac{2 F_0^2 \Omega}{m \pi v_s^2}\int_0^\pi \frac{v_g(\theta)^2 \left(\Omega - \Omega \cos (\Omega t)\cos(\omega(\theta)t)-\omega \sin(\omega(\theta)t)\sin(\Omega t)\right)}{(\Omega^2-\omega(\theta)^2)^2}\mathrm{d}\theta\\[3mm]
    \DS -\frac{ F_0^2}{m\pi v_s^2}\int_0^\pi \frac{v_g(\theta)^2\sin^2 (\Omega t)}{\Omega^2-\omega(\theta)^2}\mathrm{d}\theta.
    \end{array}
\end{equation}
The equation~(\ref{WORK_EXACT}) is the exact expression for the total energy of the semi-infinite chain. Below we investigate behavior of~$U$ at large times.

\subsection{Large time asymptotics}\label{H2}

In order to estimate large time asymptotics for the total energy, we follow the proposed in~\cite{Kuz2018} approach, simplifying expression for the energy~(\ref{WORK_EXACT}). If the frequency~$\Omega$ exceeds~$2\omega_e$, i.e., does not belong the pass-band, defined by the dispersion relation, the total energy can not be permanently supplied into the chain\footnote{Aforesaid is further shown to be valid only in the harmonic approximation.} for a reason discussed in~Sect.\ref{APP_MORE}. \edited{The energy grows until it reach some value for the time, depending on~$\Omega$, and further oscillates near it\footnote{This process is shown in~\cite{Cannas1991} for the case of kinematic loading~(see Fig.~(2)) therein. Evolution of the total energy in the case of force loading is qualitatively similar.}}. Therefore, we consider only the case where the driving frequency belongs to the pass-band~($\Omega \in (0; 2\omega_e]$). At large times, the second term in~(\ref{WORK_EXACT}) can be neglected by virtue of its boundedness, if the frequency~$\Omega$ is not about zero\footnote{Otherwise, time scale of oscillations of the total energy,~$2\pi/\Omega$,  is much greater than~$2\pi/\omega_e$ and therefore the second term in~(\ref{WORK_EXACT}) cannot be ignored. The growth of the total energy is then slower than for the situations, considered below.}. 
\par Changing the wavenumber integral to the frequency integral in first term of~(\ref{WORK_EXACT}), we rewrite expression for the energy~$U$ as
\begin{equation}\label{WORK_APPROX}
\small
    U\approx \frac{2 F_0^2 \Omega a}{m \pi v_s^2} \int_0^{2\omega_e}\frac{v^g(\omega) \left(\Omega - \Omega \cos (\Omega t)\cos(\omega t)-\omega \sin(\omega t)\sin(\Omega t)\right)}{(\Omega^2-\omega^2)^2} \mathrm{d}\omega,
    \normalsize
\end{equation}
where~$v^g$ is determined by Eq.~(\ref{VEL23_3}). It is seen from expression~(\ref{WORK_APPROX}) that the main contribution to the integral comes from \edited{the} vicinity of the point~$\omega=\Omega$. Changing the variable in the integral in~(\ref{WORK_APPROX})~$\epsilon=\Omega-\omega$ and, using the following asymptotic expansions with respect to~$\epsilon/\omega_e$:
\begin{equation}\label{777_1}
    \DS \Omega - \Omega \cos (\Omega t)\cos(\omega t)-\omega \sin(\omega t)\sin(\Omega t) = 2 \Omega \sin^2 \left(\frac{\epsilon t}{2}\right)+O\left(\frac{\epsilon}{\omega_e}  \right)
\end{equation}
\edited{and}
\begin{equation}\label{777_2}
    \Omega^2-\omega^2 = 2\Omega \epsilon+O \left( \left(\frac{\epsilon}{\omega_e} \right)^2 \right),
   \quad v^g(\omega)= v^g(\Omega)+O \left(\frac{\epsilon}{\omega_e}\right),
\end{equation}
we rewrite the  expression for the total energy as
\begin{equation}\label{PRE_FIN_ENERG}
    U\approx \frac{F_0^2 a v^g(\Omega)}{m\pi v_s^2} \int_{\Omega-2\omega_e}^\Omega \frac{\sin ^2 \left(\frac{\epsilon t}{2}\right)}{\epsilon^2}\mathrm{d}\epsilon.
\end{equation}
Tending in the integral in~(\ref{PRE_FIN_ENERG})~$\epsilon t\rightarrow \infty$ and using the identity
\begin{equation}\label{EQLTY}
    \int_0^\infty \frac{\sin ^2\frac{x}{2}}{x^2}\mathrm{d}x=\frac{\pi}{4},
\end{equation}
we write expression for energy, supplied into the chain at large times, in the final form:
\begin{equation}\label{FIN_ENERG}
     U\approx \frac{F_0^2 v^g(\Omega)a t}{2m v_s^2}.
\end{equation}
According to~(\ref{FIN_ENERG}), the total energy, supplied into the chain by periodic force loading, linearly grows in time. The reason of the growth is propagation of the non-decaying oscillations through the chain with the group velocity~(see Sect.~\ref{APP_LESS} for details). The more frequency~$\Omega$ is the slower front~$\Tilde{n}_g$, carrying the energy, propagates and thus the rate of energy supply is less. For the first time the result was obtained in~\cite{mokole1990exact} in terms of the average power supply~(i.e.,~$\dot{U}$).
\par Note that the total energy, supplied into the infinite Hooke chain and is expressed as~(see the expression~(18) in~\cite{Kuz2018})
\begin{equation}\label{ENERG_INFINITE}
    U_\mathrm{inf}\approx \frac{F_0^2 at}{4mv^g(\Omega)},
\end{equation}
grows also linearly in time but is inversely proportional to the group velocity, owing to which rate of energy grows with increasing of~$\Omega$. 
\par
\begin{figure}[htb]
\center{\includegraphics[width=0.75\linewidth]{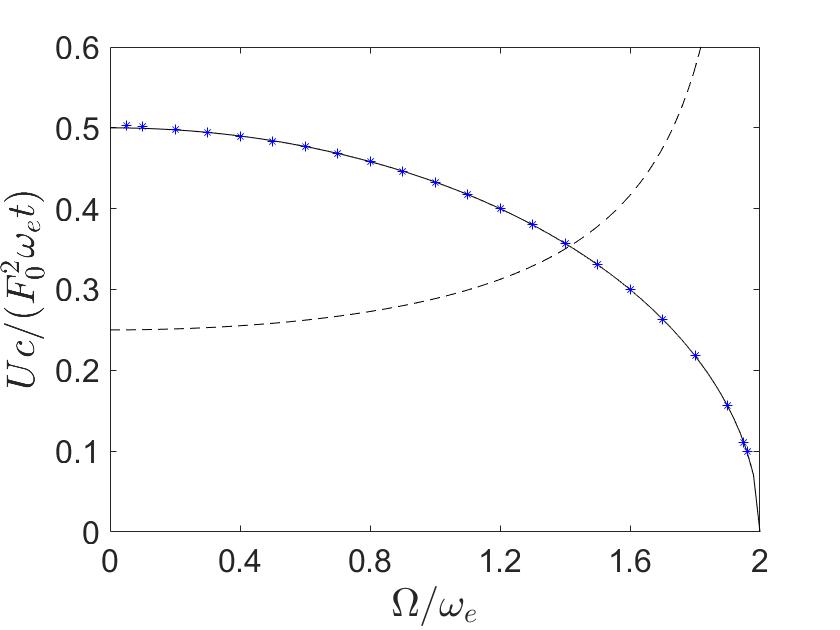}}
\caption{Dependence of the total energy \edited{growth rate} in the semi-infinite~((\ref{FIN_ENERG}), solid line) and infinite~((\ref{ENERG_INFINITE}), dashed line) chains versus the excitation frequency~$\Omega$. Results of the numerical simulations are shown by the asterisks.}
\label{fig15}
\end{figure}

In order to check the estimation~(\ref{FIN_ENERG}), we perform numerical integration of Eqs.~(\ref{EQ_harm}) in the same way as discussed in~Sect.~\ref{VEL2}. The obtained particle velocities and displacements are used for calculation the total energy of the chain. Comparison  of the expressions~(\ref{FIN_ENERG}) with the numerical solution and~(\ref{ENERG_INFINITE}) is shown in Fig.~\ref{fig15}.

It is also seen that, \edited{firstly}, asymptotic solution for the energy in the semi-infinite chain is estimated with high accuracy. \edited{Secondly, the} energy is supplied into the semi-infinite chain faster than into the infinite chain at the driving frequencies~$\Omega \in \left(0; \sqrt{2}\omega_e \right)$. At the driving frequencies~$\Omega \in \left( \sqrt{2}\omega_e; 2\omega_e \right)$ energy is supplied to the infinite chain faster than to the semi-infinite chain.  Note that at low frequencies the energy supplied into the semi-infinite chain is approximately~2 times more than the one supplied into the infinite chain. \edited{Thus, \textit{only}} for~$\Omega \ll \omega_e$  the problem of energy supply may be considered symmetric with respect to the boundary. 
\par Note also that \edited{the rate of} energy supply into the infinite elastic bar~(model of the Hooke chain in the continuum limit), caused by the force loading, is estimated in~\cite{Kuz2018}~(see~\edited{the} expression~(50) therein). The expression for the energy, \edited{supplied} to the infinite bar has the same form as for the infinite Hooke chain, if~$v^g$ in~(\ref{ENERG_INFINITE}) is replaced by~$v_s$. The latter is satisfied also for the semi-infinite elastic bar~(according to the comparison of~(\ref{FIN_ENERG}) and the expression~(49) in~\cite{Kuz2018}). Therefore,~\edited{the rate of energy supply} into the semi-infinite Hooke chain can be estimated by the continuum model only for low driving frequencies.

\par At the cut-off frequency,~$\Omega=2\omega_e$~(at zero group velocity), the asymptotic approximation~(\ref{FIN_ENERG}) becomes inapplicable. In this case, we derive the latter as follows. Rewrite the asymptotic expansions~\edited{(\ref{777_1}---\ref{777_2})} with respect to~$\epsilon=2\omega_e-\omega$ as
\begin{equation}\label{999_1}
\begin{array}{l}
    \DS \Omega - \Omega \cos (\Omega t)\cos(\omega t)-\omega \sin(\omega t)\sin(\Omega t) = 4\omega_e \sin^2 \left(\frac{\epsilon t}{2}\right)+O\left(\frac{\epsilon}{\omega_e} \right),
   \end{array}
\end{equation}
\begin{equation}\label{999_2}
     \Omega^2-\omega^2 = 4\omega_e \epsilon+O \left(\left(\frac{\epsilon}{\omega_e}  \right)^2 \right),\quad v^g\approx v_s \sqrt{\DS\frac{\epsilon}{\omega_e}}+ O \left(\left(\frac{\epsilon}{\omega_e}  \right)^\frac{3}{2}\right).
\end{equation}
Therefore, expression for the energy can be approximately rewritten in the form
\begin{equation}\label{U1_MAX}
     U\approx \frac{F_0^2a\sqrt{t}}{\pi mv_s\sqrt{\omega_e}}\int_0^{2\omega_e t} \frac{\sin ^2 \left(\frac{\epsilon t}{2}\right)}{(\epsilon t)^\frac{3}{2}}\mathrm{d}(\epsilon t).
\end{equation}
Tending in~(\ref{U1_MAX})~$\epsilon t \rightarrow \infty$ and using the identity
\begin{equation}\label{EQLTY_2}
    \int_0^\infty \frac{\sin ^2 \frac{x}{2}}{x^\frac{3}{2}}\mathrm{d}x=\sqrt{\frac{\pi}{2}},
\end{equation}
we write expression for energy, supplied into the chain at large times, in the final form:

\begin{equation}\label{U_MAX}
      U\approx \frac{F_0^2 \sqrt{t}}{m\sqrt{2\pi \omega_e^3}}.
\end{equation}
Therefore, at the frequency~$\Omega=2\omega_e$ \edited{the} energy grows in time proportionally to~$\sqrt{t}$. In this case, the reason of the growth is increasing in time of the field of particle velocities~(and, therefore, field of particle displacements). Note that \edited{the} energy, supplied into the infinite chain at the same frequency,  grows in time as~$t^{3/2}$~(see expression~(19) in~\cite{Kuz2018}) and the growth is caused by the same~\cite{Hemmer}. The reason of different asymptotic behavior of the total energy of the infinite and semi-infinite chains \edited{is, apparently, peculiarities of the fields of particle velocity. Namely, in contrast to the semi-infinite chain, the expression for particle velocity in the infinite Hooke chain does not decrease with increasing~$n$~(see time derivative of Eq.(1.58) in~\cite{Hemmer})}. A comparison of the expression~(\ref{U_MAX}) with the numerical results is discussed in Sect.~\ref{AH2}. 

Thus, analytical estimation of the energy transfer into the semi-infinite
Hooke chain is performed. We show in the next section that the expression~(\ref{FIN_ENERG}) serves for zeroth approximation for the solution in the anharmonic case.

\section{Energy transfer into the~$\beta$-FPUT chain}\label{AH}

In the section, we investigate manifestation of anharmonicity on the behavior of \edited{the} energy, supplied into the semi-infinite chain and perform large-time asymptotic estimation for it. The total energy can be calculated via Eq.~(\ref{WORK}) by substitution of asymptotic expansion for the velocity of zeroth particle up to order of~$\beta F_0^2/c^3$, which can be obtained by means of the perturbation analysis~(see, e.g., \cite{Narisetti, Sepehri}). However, this approach involves technical difficulties and, therefore, we propose the following strategy below.

\subsection{Quasiharmonic approximation for the energy}\label{AH1}
We refer to the wave turbulence theory~(see, e.g.,~\cite{Kolm1992}). According to it, the main contribution to the energy exchange between waves, induced by the cubic nonlinearity, comes from the four-wave trivial resonant~(2 $\rightarrow$ 2)-type interactions. The dispersion relation, corresponding to the harmonic part of the Hamiltonian is the renormalized dispersion relation\footnote{The renormalized dispersion relation may correspond  either to expansion spectrum~(for the $\beta$-FPUT chain, see below) or to the shifted one~(for the nonlinear Klein-Gordon equation, see~\cite{Shirokov}).}. For the~$\beta$-FPUT chain, the latter is analyzed in~\cite{Gershgorin2005, Gershgorin2007,LeeW}. In order to estimate the total energy, we use the expression for the renormalized dispersion relation, obtained in the paper~\cite{Gershgorin2005}~(see Eq.~(3) therein).\par
\small
    \textit{Remark}~1. The expression for the renormalized dispersion relation of~$\beta$-FPUT chain may also be derived in different ways, for instance, as discussed in~\cite{Gershgorin2007}.  Comparison of these ways with one, considered here, is beyond the scope of present paper.\\[2mm]
\normalsize
For the semi-infinite chain, we rewrite equation for the latter as
\begin{equation}\label{RenD}
    \Tilde{\omega}=\lambda \omega, \quad \lambda = \sqrt{1+\frac{3\beta}{2\pi c\omega_e^2}\int_{-\pi}^{\pi} \langle \hat{u}(t)^2\rangle_t 
 \omega(\theta)^2\mathrm{d}\theta}.
\end{equation}
where~$\lambda$ is the renormalized factor; $\hat{u}$ is the direct DCT of the particle displacement, which, in general case~(for arbitrary~$\beta$), is unknown;
the $\langle ... \rangle_t$ stands for the time averaging. Here, we average over period of the weakly anharmonic Duffing oscillator\footnote{See Sect. 4.2.1 in~\cite{Kovacic} for details. Expansion of Eq.~(4.2.15) with respect to the nonlinear parameter yields equation for the period in the form, represented in Eq.(\ref{DuffT}).}:
\begin{equation}\label{DuffT}
    \langle ... \rangle_t=\frac{1}{\tau_D}\int_0^{\tau_D} ... \mathrm{d}t,\quad \tau_D=\frac{2\pi}{\omega_e}\left(1-\frac{3\beta F_0^2}{8c^3}\right)+O\left(\frac{\beta^2 F_0^4}{c^6}\right).
\end{equation}
\small \par
\begin{figure}[htb]
\center{\includegraphics[width=0.75\linewidth]{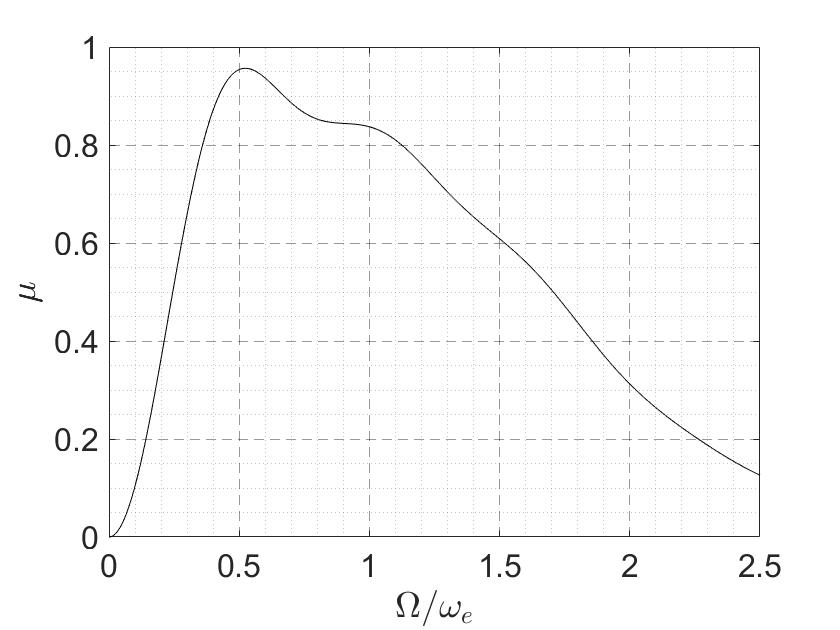}}
\caption{Dependence of the function~$\mu$ on the \edited{driving} frequency~$\Omega$.}
\label{fig4}
\end{figure}

\textit{Remark}~2. The renormalized dispersion relation is derived in~\cite{Gershgorin2005} under assumption that the $\beta$-FPUT chain is strongly nonlinear and is in thermal equilibrium. This is not our case. Here,~$\Tilde{\omega}$ is understood as some equivalent linear dispersion relation, which is analogous to the equivalent linear frequency, in terms of which some dynamical problems of nonlinear \edited{oscillations} may be solved~(see, e.g.,~\cite{Panovko,HB}).
\normalsize
\newline
Note that, firstly, the renormalized factor does not depend on the wavenumber~$\theta$. Secondly, since we calculate~$\lambda$ with accuracy up to order of~$\beta F_0^2/c^3$, the expression for~$\hat{u}$ in the harmonic approximation can be used. Substitution of~(\ref{IMAG1}) at~$\eta=0$ and~(\ref{DuffT}) to~(\ref{RenD}) and further simplification with the remaining terms of order of $\beta F_0^2/c^3$, yields the following approximation for~$\lambda$:
\begin{equation}\label{renFACT_1}
    \lambda \approx 1\DS+\frac{\beta F_0^2}{c^3}\mu(\Omega),
\end{equation}
\begin{equation}\label{renFACT_2}
    \mu(\Omega)\defeq\DS\frac{3\omega_e^2}{8\pi^2}\int_0^{2\pi} \int_0^{2\pi} \frac{\cos^2\frac{\theta}{2}\Bigl(\Omega \sin(\omega(\theta) t)-\omega(\theta) \sin(\Omega t) \Bigr)^2}{(\Omega^2-\omega(\theta)^2)^2}  \mathrm{d}({\omega_e t})\mathrm{d}\theta. 
\end{equation}
In simulations, the integral in~\edited{(\ref{renFACT_2})} is calculated using the rectangular rule with division of the integration domain on~$10^6$ equal squares. Dependence of the function~$\mu$ determining the renormalization factor on the excitation frequency is shown in Fig.~\ref{fig4}. 
 For convenience, we rewrite~$\mu$ in the following approximate form, which  is satisfied for~$\Omega/\omega_e \in [1.5; 2.1]$\footnote{Choice of range of the driving frequencies is explained in Sect.\ref{AH2}.}:
\begin{equation}\label{FIT}
    \mu \approx 0.848 \left(\frac{\Omega}{\omega_e}\right)^3-4.580\left(\frac{\Omega}{\omega_e}\right)^2+7.591\left(\frac{\Omega}{\omega_e}\right)-3.332.
\end{equation}
\par
We assume that the main contribution to energy transmission comes from the oscillations with frequencies, which belong to the renormalized dispersion relation~$\Tilde{\omega}$ and from the harmonic oscillations with the driving frequency~$\Omega$~(subharmonic~(with the frequencies~$\kappa$ times less~$\Omega$~($\kappa\in \mathbb{N}, \kappa>1$), see, e.g.,~\cite{KeenB} for details) and superharmonic~(with the frequencies~$m$ times more~$\Omega$~($\kappa\in \mathbb{N}, \kappa>1$)) oscillations are neglected). Then, taking into account the first equation in Eqs.~(\ref{RenD}),~(\ref{renFACT_1}) and~(\ref{renFACT_2}) and supposing that~$\Omega$ is not too close to zero, we rewrite Eq.~(\ref{WORK_APPROX}) for the energy, transmitted into the chain, with respect to~$\Tilde{\omega}$, namely 
\begin{equation}\label{WORK_NONL}
    U\approx \frac{2 F_0^2 \Omega a}{m \lambda \pi v_s^2} \int_0^{\Tilde{\omega}_{\max}}\frac{v^g \left(\frac{\Tilde{\omega}}{\lambda}\right) \left(\Omega - \Omega \cos (\Omega t)\cos(\Tilde{\omega} t)-\Tilde{\omega} \sin(\Tilde{\omega}t)\sin(\Omega t)\right)}{(\Omega^2-\Tilde{\omega}^2)^2} \mathrm{d}\Tilde{\omega},
\end{equation}
where~$\Tilde{\omega}_{\max}$ is the maximal value of the renormalized dispersion relation. Performing the same operations as done in Sect.~\ref{H2} yields the following asymptotic approximation for the total energy:
\begin{equation}\label{FIN_ENERG_NONL}
    U\approx \DS\frac{F_0^2 v^g \left(\DS\frac{\Omega}{\lambda} \right)a t}{2m \lambda v_s^2}.
\end{equation}
Substituting~(\ref{renFACT_1}) and (\ref{renFACT_2}) to~(\ref{FIN_ENERG_NONL}) and preserving the terms of order of~$\beta F_0^2/c^3$, we obtain the final expression for the total energy

\begin{equation}\label{ENERG_NONL}
    U\approx \DS\frac{F_0^2 v^g \left(\Tilde{\Omega}\right)a t}{2m  v_s^2}\left(1-\frac{\beta F_0^2}{c^3} \mu \right), \quad \Tilde{\Omega}=\Omega \left(1-\frac{\beta F_0^2}{c^3} \mu\right).
\end{equation}
Therefore, expression for the total energy, supplied into the chain, is represented as a sum of harmonic approximation at the  frequency~$\Tilde{\Omega}$ and correction, directly proportional to the first term. Further, we refer~(\ref{ENERG_NONL}) to as a \textit{quasiharmonic approximation}.
It is seen from~(\ref{ENERG_NONL}) that, firstly, as expected, energy dependence on~$F_0^2$ is not linear. Secondly, the expression~(\ref{ENERG_NONL}) is valid not only for the cutoff frequency of the Hooke chain~($\Omega=2\omega_e$) but for frequencies, which may slightly~\textit{exceed} this value. From~(\ref{ENERG_NONL}), we determine a maximal frequency, permitting passing waves into the chain,~$\Omega_\mathrm{cr}$, satisfies the equation 
\begin{equation}\label{EQ_MAX}
\Omega_\mathrm{cr}\left(1-\frac{\beta F_0^2}{c^3} \mu(\Omega_\mathrm{cr})\right)=2\omega_e.
\end{equation}
 The value of~$\Omega_\mathrm{cr}$ is found from Eq.(\ref{EQ_MAX}) with accuracy up to order of~$\beta F_0^2/c^3$ as~(see Appendix~\ref{DDD}):
\begin{equation}\label{APPR}
\Omega_{\mathrm{cr}}\approx\left(2+2\mu(2\omega_e)\frac{\beta F_0^2}{c^3}\right)\omega_e.
\end{equation}
From~(\ref{FIT}), $\mu(2\omega_e)\approx 0.314$.  In order to check accuracy of the approximations~(\ref{ENERG_NONL}) and (\ref{APPR}), we perform the numerical simulations by integration of Eqs.~(\ref{DynEQ})~\edited{the initial conditions~(\ref{IC})} in the same way as described in~Sect.~\ref{VEL2}. The obtained particle velocities and displacements are used for calculation the total energy of the chain.  Further, we investigate evolution of the total energy at large times and for different types of the driving frequencies.
 
\subsection{Energy propagation at frequencies, lying in the pass-band of the Hooke chain} \label{AH2}
\begin{figure}[htb]
\begin{minipage}{0.5\linewidth}
\center{\includegraphics[width=0.95\linewidth]{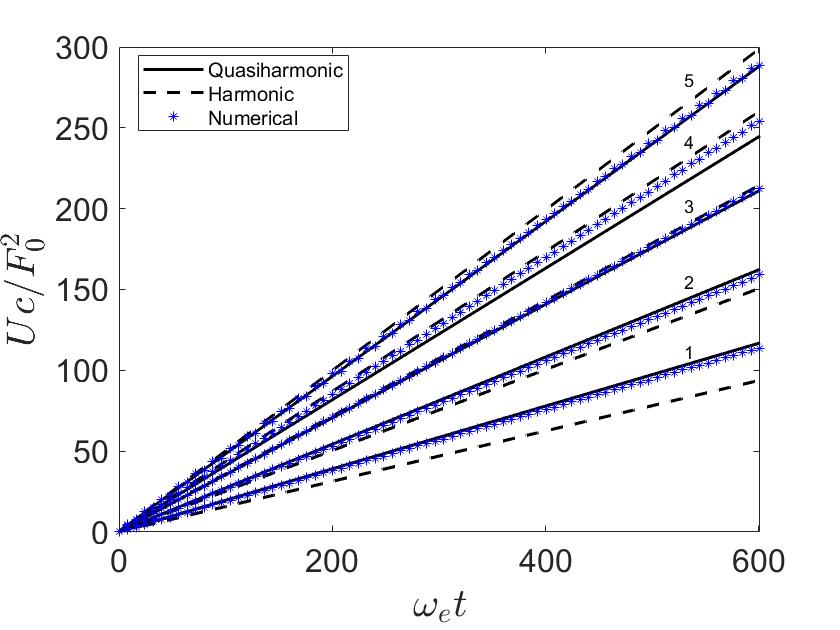} \\A }
\end{minipage}
\begin{minipage}{0.5\linewidth}
\center{\includegraphics[width=0.95\linewidth]{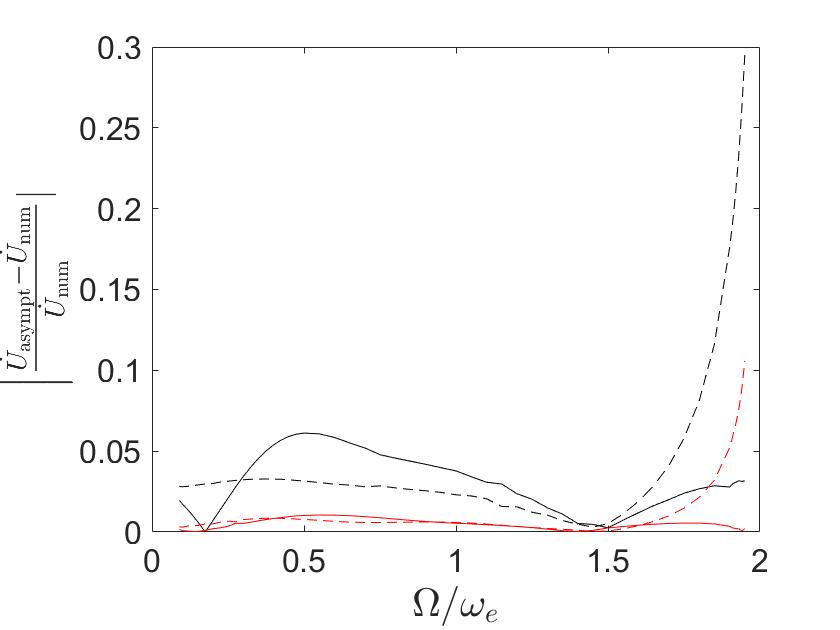} \\B}
\end{minipage}
\caption{A: Asymptotic solutions for the total energy, supplied into the semi-infinite chain for~$\beta F_0^2/c^3=0.1$ at frequencies~$\Omega=1.90\omega_e$~(1), $\Omega=1.73\omega_e$~(2), $\Omega=1.41\omega_e$~(3), $\Omega=\omega_e$~(4), $\Omega=0.20\omega_e$~(5) in the harmonic and quasiharmonic approximations~(expressions~(\ref{FIN_ENERG}) and~(\ref{ENERG_NONL}) respectively). B: Relative error of asymptotic solution for the rate of energy supply into the harmonic~(dashed line) and quasiharmonic~(solid line) approximations for~$\beta F_0^2/c^3=0.02$~(red line) and~$\beta F_0^2/c^3=0.1$~(black line).}
\label{fig5}
\end{figure}

Evolution of the total energy at large times and at~$\Omega\in (0; 2\omega_e)$ for~$\beta F_0^2/c^3=0.1$ is shown in Fig.~\ref{fig5}A.

It is seen that the asymptotics for the energy in the harmonic approximation~(\ref{FIN_ENERG}) is valid at short times for all range of the considered driving frequencies. For the frequency~$\Omega=1.41 \omega_e$, results, obtained in harmonic approximation, coincide with results, obtained in the quasiharmonic approximation~(dashed and solid lines coincide, see Fig.~\ref{fig5}A), and also with the results of numerical simulations. However, at large times, manifestation of nonlinearity occurs. For instance, for the frequencies~$\Omega=1.90\omega_e$, $\Omega=1.73\omega_e$ and~$\Omega=0.20\omega_e$ description of the energy supply is more accurate in quasiharmonic approximation than in the harmonic one. Wherein, however, for~$\Omega=\omega_e$, the harmonic approximation is more correct than the quasiharmonic one.  In order to analyze accuracy of the asymptotic approximations in detail, we calculate the relative error between~$\dot{U}_{\mathrm{asympt}}$~(using the expressions~(\ref{FIN_ENERG}) and~(\ref{ENERG_NONL}) for the harmonic and quasiharmonic approximations respectively) and  $\dot{U}_{\mathrm{num}}$, averaged over time interval, equal to~$10^3/\omega_e$, for range of the frequencies~$\Omega \in [0.09\omega_e; 1.95\omega_e]$. One can see in Fig.~\ref{fig5}B that, for the driving frequencies, which are close to the cut-off frequency of the Hooke chain, the relative error of the harmonic approximation is sufficiently higher than one of the quasiharmonic approximation. Therefore, for the driving frequency~$\Omega \lessapprox 1.5\omega_e$, energy \edited{supply} can be estimated in the harmonic approximation with reasonable accuracy. 

\begin{figure}[htb]
\begin{minipage}{0.52\linewidth}
\center{\includegraphics[width=0.9\linewidth]{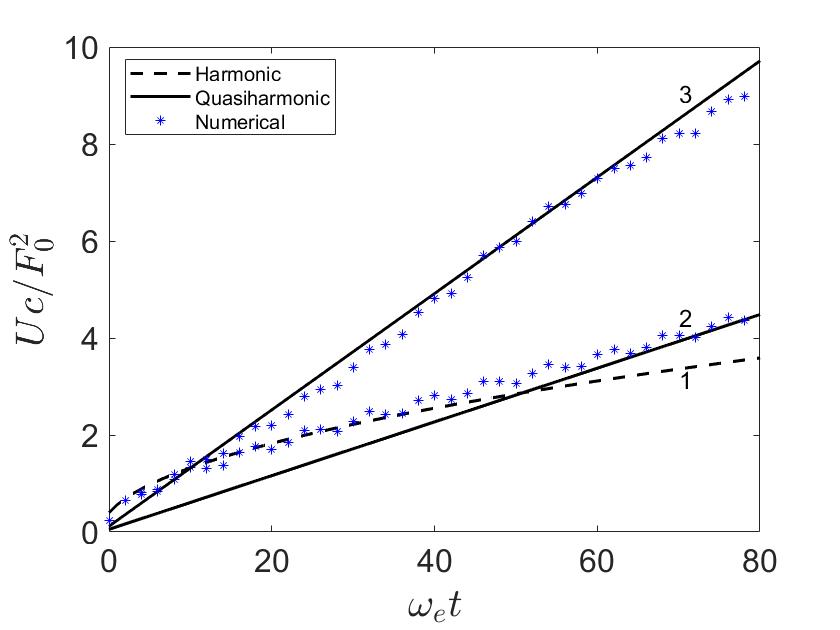}}
\end{minipage}
\begin{minipage}{0.52\linewidth}
\center{\includegraphics[width=0.9\linewidth]{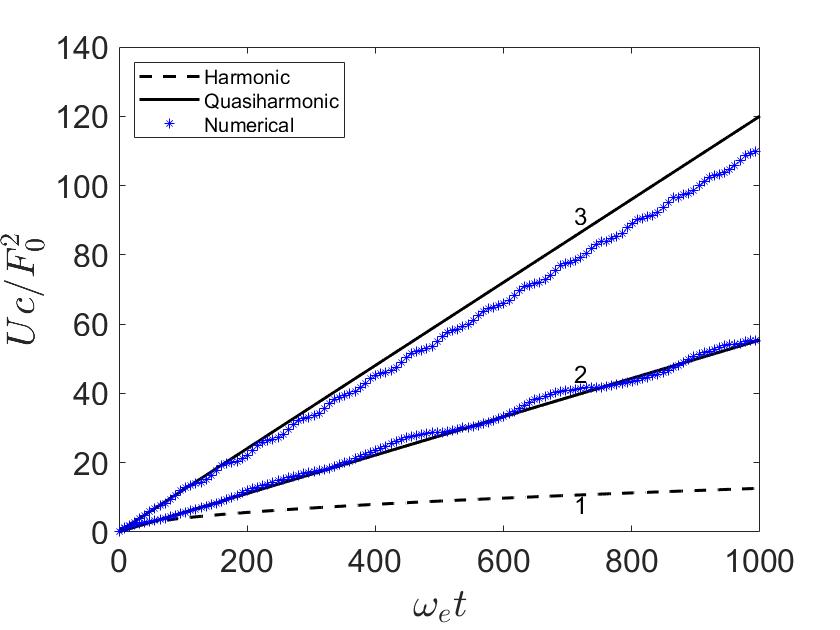}}
\end{minipage}
\caption{Evolution of energy, supplied into the semi-infinite chain at the cut-off frequency of the harmonic chain~($\Omega=2\omega_e$) at short times~(left) and at large times~(right). \edited{The} asymptotic solution is shown for $\beta F_0^2/c^3=0$~(1, expression~(\ref{U_MAX})), $\beta F_0^2/c^3=0.02$~(2, expression~(\ref{ENERG_NONL})), $\beta F_0^2/c^3=0.1$~(3, expression~(\ref{ENERG_NONL})).}
\label{fig7}
\end{figure}

Consider the evolution of the total energy, transferred into the chain at the cut-off frequency of the Hooke chain.  The time dependence of the total energy is shown in Fig.~\ref{fig7}.

It is seen in Fig.~\ref{fig7} that energy supply at~$\Omega=2\omega_e$ has two time scales. The first time scale corresponds to growth of the total energy, which can be adequately described in the harmonic approximation~(expression~(\ref{U_MAX})). Starting from some time, the rate of propagation undergoes sharp increase.
Evolution of the energy may be then described in the quasiharmonic approximation with rather high accuracy~(at least at times~$\omega_et<250$, see Fig.~\ref{fig7}~(right)).  

\subsection{Energy propagation at frequencies, lying in the stop-band of the Hooke chain} \label{AH3}

\begin{figure}[htb]
\center{\includegraphics[width=0.75\linewidth]{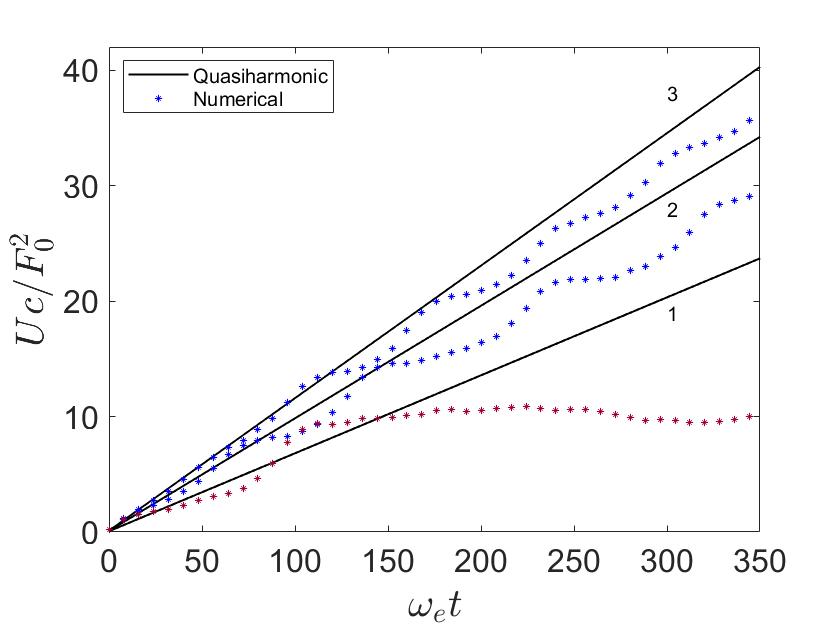}}
\caption{\edited{The} asymptotic approximation~(\ref{ENERG_NONL}) for \edited{the} energy, supplied into the semi-infinite chain at the frequencies~$\Omega=2.005\omega_e$~(3), $\Omega=2.02\omega_e$~(2), $\Omega=2.04\omega_e$~(1). Comparison of the asymptotic and numerical solutions is shown for~$\beta F_0^2/c^3=0.1$.}
\label{fig9}
\end{figure}

We consider the energy supply into the semi-infinite chain at~$\Omega > 2\omega_e$. The loading with the frequency, exceeding~$2\omega_e$, results in energy supply into the chain~(see Fig.~\ref{fig9}).  \par

In order to understand nature of sharp increase of the supplied energy, we perform set of numerical experiments~(See Sect.~\ref{AH1} for details). These experiments revealed a localized mode, which moves slower than the decaying perturbations, moving with the speed of sound. Moreover, a numerical experiment, performed for the free semi-infinite 
 Hooked chain~($\beta=0, F=0$) showed that the launched same localized mode is extinguished. According to the aforesaid observations, we suppose that the reason of the energy supply is the supratransmission and the source of the transferred into the chain energy~(as well as for~$\Omega=2\omega_e$) is, apparently, intrinsic localized modes. It is seen in Fig.~\ref{fig9} that, firstly, evolution of the total energy can be well-described in the quasiharmonic approximation~(\ref{FIN_ENERG_NONL}) at least at times~$\omega_et<150$. However, the quasiharmonic approximation becomes hereafter no longer valid owing to the either slowdown or stop of this process, therefore, at high driving frequencies and sufficiently large times, the energy does not grow linearly. Apparently, corrections of order of~$\beta^2 F_0^4/c^6$ are required for more accurate description of this process. Secondly, one can see that a small change of the driving frequency results in sufficient change of rate of the transfer.

\begin{figure}[htb]%
\center{\includegraphics[scale=0.6]{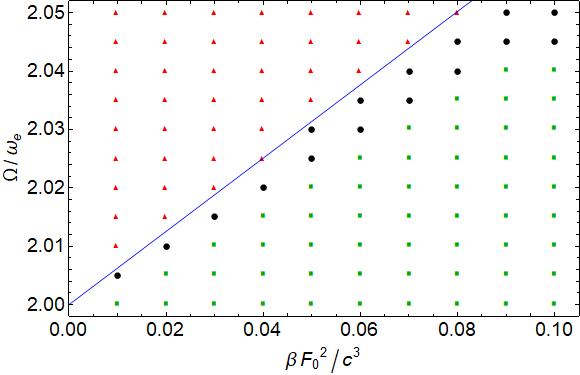}}
\caption{Diagram of \edited{relations} between anharmonic force~$\beta F_0^2/c^3$ and~$\Omega/\omega_e$, corresponding to \edited{ability} of the energy {transmission}. \edited{The quasiharmonic approximation for the maximal frequency, permitting the transmission~((\ref{APPR}), solid line) and the numerical results~(triangles, circles and squares) are shown.}  \edited{Triangles and circles denote the relations, corresponding to inability of the energy transmission, Squares denote the relations, corresponding to ability of the energy transmission.} }
\label{fig10}
\end{figure}

 Consider the case when the excitation frequency is close to~$\Omega_\mathrm{cr}$. We calculate numerically the total energy of the chain for frequencies~$2\omega_e \leqslant \Omega \leqslant \Omega_\mathrm{cr}$ and~$0<\beta F_0^2/c^3 \leqslant 0.1$ and then compare the results with the quasiharmonic approximation~(\ref{ENERG_NONL}).
The relations between~$\Omega/\omega_e$ and~$\beta F_0^2/c^3$, corresponding to the different situations, concerning energy transmission into the chain, are shown in~Fig.~\ref{fig10}. 
There are such relations, at which the energy is not transmitted into the chain, according to both simulations and the quasiharmonic approximation~(\ref{ENERG_NONL}). These relations are marked in the diagram, shown in~Fig.~\ref{fig10} by the~\textit{triangles} above the dividing line, determining~$\Omega_{\mathrm{cr}}$ by means of the prediction~(\ref{APPR}). 
There are such values of $\Omega/\omega_e$ and~$\beta F_0^2/c^3$, at which energy is transmitted into the chain and the quasiharmonic approximation is in agreement with it~(marked by the \textit{squares}). However, according to the numerical results, several values of $\Omega/\omega_e$ and~$\beta F_0^2/c^3$ correspond to the inability to pass energy into the chain, while the quasiharmonic approximation~(\ref{APPR}) shows the contrary. These values are marked by the \textit{circles} and are distributed between the triangles and squares. Therefore, the real value of the frequency, permitting the energy transmission, is less than~$\Omega_{\mathrm{cr}}$. Apparently, neglecting 
non-trivial resonant or~($3\rightarrow 1$)- and~($4\rightarrow 0$)-type interactions in the \edited{estimation} of the total energy might result in some error in~(\ref{WORK_NONL}) and~(\ref{APPR}). We do not consider this question in the current work in detail.
\par The expression~(\ref{APPR}) may be estimated as an upper boundary of the critical frequency. The lower boundary of the critical frequencies can be contingently determined by a straight line to the point~$(0.10; 2.045)$. The slope of the latter~($0.450$) is less than one, corresponding to the upper boundary~($2\mu(2\omega_e)\approx0.628$).

\small \par
\textit{Remark}~3. We have reproduced the numerical calculations of the total energy, having replaced~$\sin(\Omega t)$ in right part of Eq.~(\ref{DynEQ}) by~$\cos(\Omega t)$ and have convinced that this action does not affect~\edited{qualitatively} on the obtained in this section~(and in the present paper at whole) results. 
\normalsize

\section{Conclusions}\label{Concl}
Energy supply into the semi-infinite $\beta$-FPUT chain by the sudden periodic force loading was considered.  Asymptotic peculiarities of the growth of the energy supplied at large times have been estimated. 

A modification of the Hemmer approach, which allows to obtain an approximate solution of a dynamical problem in the harmonic approximation~(namely, for the Hooke chain) is proposed~(Sect.\ref{VEL2}). Herewith, the contribution, corresponding to the forced oscillations, is found exactly while for the second contribution, corresponding tho the natural oscillations, the large-time asymptotic estimation is performed. Obtained expressions for the contributions are used to construct the approximate solution, which is a refined version of the dynamical problem solution obtained in~\cite{mokole1990exact} and upon which the explanation of mechanism of energy supply into the chain at large time is based. At low driving frequencies, an effect of \edited{the dispersion may be shown} to be neglected: a process of energy supply is qualitatively indistinguishable from the same process both in the infinite chain and in the elastic rod~(see~\cite{Kuz2018}). However, \edited{the} rate of the energy supply decreases with increasing the driving frequency~(in contrast to the infinite chain) \edited{and the problem of energy supply into the semi-infinite chain, in general, can not be considered symmetric with respect to the boundary.}

In the weakly anharmonic case, the process of energy supply can be described in the harmonic approximation for short times at driving frequencies, lying in the pass-band of the Hooke chain. At large times, the harmonic approximation is reasonably accurate for~$\Omega \lessapprox 1.5\omega_e$. For the general case, an asymptotic solution for the total energy~(referred further to as the quasiharmonic approximation) was obtained~(expression~(\ref{ENERG_NONL})) by dint of the derived in~\cite{Gershgorin2005} and rewritten for the semi-infinite chain equation for the renormalized dispersion relation. \edited{The quasiharmonic approximation was shown to be suitable} for the description of energy supply at large times~(of order at least 
 of~$150/\omega_e$) under force loading with the driving frequency, which both may be equal to~$2\omega_e$ and may lie in the stop-band of the Hooke chain.  
By using the first term in~(\ref{ENERG_NONL}), approximation for the frequency, above which the energy transmission into the chain is impossible~($\Omega_\mathrm{cr}$), was obtained~(expression~(\ref{APPR})). However, the results of numerical analysis have shown that the real value of this frequency is less than~$\Omega_\mathrm{cr}$ for any power of anharmonicity. 

The results of the current work can be generalized to the cases of the interactions via more physically realistic~(than~$\beta$-FPUT) potentials. Consideration of the cases of loading by a force, arbitrary changing in time, will be also an interesting extension of the present paper. The results with regard to~\edited{the} energy supply in this case may serve for development of theory of heat transport at the nanoscale in the frameworks of the lattice dynamics approach.  In particular, a problem of propagation of thermal wave through the interface between two dissimilar semi-infinite homogeneous chains can be interpreted in terms of dynamics of the one chain segment, undergone an equivalent force, acting from the second segment. An expression for this force is unknown. However, using approaches based on the energy dynamics~\cite{ZAMM}, one can obtain some quantities without calculation of the equivalent force itself~(see~\cite{Kuz2023}). The results concerning asymptotic estimates of the supplied total energy are expected to serve for simplification of \edited{the} aforesaid problem, concerning inhomogeneous anharmonic lattices. Understanding of the heat transport therein is important for development of thermal diode, \edited{thermal transistor and heat pump} techniques~(see, e.g.,~\cite{Terraneo, Kobayashi, Li2012, Malik, ai2010heat}).

\section*{Acknowledgements}
The work is supported by the Russian Science Foundation~(Grant No.\;22-11-00338). The author is deeply grateful to S.A. Shcherbinin, S.N. Gavrilov, V.A. Kuzkin, A.M. Krivtsov, \edited{E.A. Korznikova} and S.V. Dmitriev for useful and \edited{stimulating} discussions \edited{and to anonymous referees for the valuable comments}. The work is dedicated to the memory of Professor Dmitry Anatolyevich Indeitsev.

\section*{Conflict of interest}
None to declare. 

\begin{appendices}
   \section{Derivation of Eq.~(\ref{ForcedContr})}\label{AAA}
   Here, we derive Eq. for the contribution~$v_{n}^\Omega$ in the closed form. Using the following approximations due to $\eta/\omega_e$ is infinitesimal: 
\begin{equation}\label{A_0}
    \begin{array}{l}
     \DS \phi_2 \approx \left[ 
      \begin{gathered} 
        \pi, \, \omega < \Omega; \\ 
        0,\, \omega > \Omega, \\ 
      \end{gathered} 
\right. \quad \sqrt{\Omega^2-\omega^2\pm 2\mathrm{i}\eta \Omega}\approx \sqrt{\Omega^2-\omega^2}\pm \frac{\mathrm{i}\eta \Omega}{\sqrt{\Omega^2-\omega^2}},
    \end{array}
\end{equation}

rewrite the expression for~$v_{n}^\Omega$ in Eq.~(\ref{Vel3}) as follows:
\begin{equation}\label{A_1}
    v_{n}^\Omega\approx -\frac{F_0\Omega \cos(\Omega t)H(t)}{m}\left(\mathrm{I}_{n}+\mathrm{I}_{n+1}+\mathrm{c.c.}\right),\quad \mathrm{I}_{n}=\frac{1}{4\pi}\int_{-\pi}^{\pi}\frac{e^{\mathrm{i}n\theta}\mathrm{d}\theta}{\Omega^2-\omega^2-\mathrm{i}\eta \Omega},
\end{equation}
where~$\mathrm{c.c.}$ stands for the complex conjugate terms. 
We transform the integral in Eq.~(\ref{A_1}) to the unit circle integral. Put~$z=e^{\mathrm{i}\theta}$. Then
\begin{equation}\label{A_3}
    \mathrm{d}\theta=-\mathrm{i}\frac{\mathrm{d}z}{z}, \qquad 2\cos \theta=z+\frac{1}{z},
\end{equation}
and, therefore,
\begin{equation}\label{A_4}
    \mathrm{I}_n=-\frac{\mathrm{i}}{4 \pi}\oint_{\vert z\vert=1}\frac{z^{n}\mathrm{d}z}{\omega_e^2z^2+(\Omega^2-2\omega_e^2-\mathrm{i}\eta \Omega)z+\omega_e^2
    }.
\end{equation}
Rewrite Eq.~(\ref{A_4}) as
\begin{equation}\label{A_5}
    \mathrm{I}_n=-\frac{\mathrm{i}}{4\pi \omega_e^2(z_1-z_2)}\left(\oint_{\vert z\vert=1}\frac{z^n\mathrm{d}z}{z-z_1}-\oint_{\vert z\vert=1}\frac{z^n\mathrm{d}z}{z-z_2}\right),
\end{equation}
\begin{equation}\label{A_5_5}
     z_{1,2}=\frac{2\omega_e^2-\Omega^2+\mathrm{i}\eta\Omega\pm \mathrm{i}\sqrt{\Omega}\sqrt{\Omega-\mathrm{i}\eta}\sqrt{4\omega_e^2-\Omega^2+\mathrm{i}\eta \Omega}}{2\omega_e^2}.
\end{equation}
where~$z_1$ and~$z_2$ are roots of the denominator of the integrand in Eq.~(\ref{A_4}), i.e., poles. Rewrite the latter in the limit~$\eta\rightarrow 0+$:
\begin{equation}\label{A_6}
    z_{1,2}=1-\frac{\Omega^2}{2\omega_e^2}\pm\mathrm{i}\left(\frac{\Omega\sqrt{4\omega_e^2-\Omega^2}}{2\omega_e^2}\pm0\right).
\end{equation}
\par Consider~$0<\Omega<2\omega_e$. In this case, the pole~$z_2$ lies inside the unit circle~$\vert z\vert<1$ only. Therefore, we take into account the residue at~$z_2$ only and the first term in~(\ref{A_5}) equals zero. Therefore, we have
\begin{equation}\label{A_7}
    \mathrm{I}_{n}=\frac{\mathrm{i}z_2^n}{2\Omega \sqrt{4\omega_e^2-\Omega^2}}=\frac{(-1)^ne^{2\mathrm{i}\varphi n}}{-2\mathrm{i}\Omega\sqrt{4\omega_e^2-\Omega^2}},\quad \varphi=\arccos{\frac{\Omega}{2\omega_e}}.
\end{equation}
\par Consider~$\Omega>2\omega_e$. In this case, Eqs. for~$z_{1,2}$ have the form
\begin{equation}\label{A_8}
    z_{1,2}=1-\frac{\Omega^2\pm \Omega\sqrt{\Omega^2-4\omega_e^2}}{2\omega_e^2}.
\end{equation}
Substituting~$z_{1,2}$ to Eq.~(\ref{A_5}) with taking into account~$\vert z_2 \vert<1$ yields 
\begin{equation}\label{A_9}
\begin{array}{l}
    \DS \mathrm{I}_n=\frac{z_2^n}{2\Omega\sqrt{\Omega^2-4\omega_e^2}}=\frac{(-1)^ne^{-\gamma n}}{2\Omega\sqrt{\Omega^2-4\omega_e^2}},\\[2mm]
    \DS \gamma=-\ln \left(\frac{\Omega^2-\Omega\sqrt{\Omega^2-4\omega_e^2}}{2\omega_e^2}-1\right)=2\mathrm{arccosh}\frac{\Omega}{2\omega_e}.
    \end{array}
\end{equation}

\par Consider~$\Omega=2\omega_e$. The integral for~$\mathrm{I}_n$ and its complex conjugation,~$\bar{\mathrm{I}}_n$, determined in~(\ref{A_1}), can be analogously transformed into the unit circle integral and then to the form
\begin{equation}\label{A_10}
\begin{array}{l}
    \DS \mathrm{I}_n=-\frac{\mathrm{i}}{4\pi \omega_e^2}\oint_{\vert z\vert=1}\frac{z^n\mathrm{d}z}{(z+1-\mathrm{i}0)^2}=0, 
\end{array}
\end{equation}
\begin{equation}\label{A_10_5}
    \bar{\mathrm{I}}_n=-\frac{\mathrm{i}}{4\pi \omega_e^2}\oint_{\vert z\vert=1}\frac{z^{-n}\mathrm{d}z}{(z+1+\mathrm{i}0)^2}=\frac{(-1)^{(n+1)}n}{2\omega_e^2}.
\end{equation}
Here, the following identities~\cite{GelfandShilov}
\begin{equation}\label{A_11}
   \frac{1}{(z\pm \mathrm{i}0)^2}=\pm \mathrm{i}\pi\delta'(z)+\mathrm{p.v.}\frac{1}{z^2},
\end{equation}
where~$\mathrm{p.v.}$ stands for the Cauchy principle value, are used.  
Substitution of Eqs.~(\ref{A_7}), (\ref{A_9}), (\ref{A_10}), (\ref{A_10_5}) to Eq.~(\ref{A_1}) with respect to the corresponding values of the driving frequency and with further simplifications yields Eq.(\ref{ForcedContr}). Note that the expressions~(\ref{A_7}), (\ref{A_9}), (\ref{A_10_5}) were previously obtained in~\cite{Hemmer}. However, Eqs.~(\ref{A_7})+$\mathrm{c.c.}$ and~(\ref{A_10}) were obtained in the sense of the Cauchy principal value only, while, here, we do this exactly by using the limiting absorption principle.

\section{Derivation of the expression~(\ref{CritEq})}\label{BBB}

Rewrite Eq. for~$v_{n}^\omega$ in~Eq.~(\ref{Vel2}) as follows:
\begin{equation}\label{VEL21_1}
    v_{n}^\omega=\frac{2F_0\Omega e^{-\eta t} H(t)}{m\pi}\int_0^{\pi}\frac{(\Omega^2-\omega(\theta)^2)\cos (\omega(\theta) t)}{(\Omega^2-\omega(\theta)^2)^2+4\eta^2\Omega^2}\cos\frac{(2n+1)\theta}{2} \cos \frac{\theta}{2} \mathrm{d}\theta.
\end{equation}
Here, we use the following approximations due to $\eta/\omega_e$ is infinitesimal:
\begin{equation}\label{VEL21_011}
\begin{array}{l}
   \DS \phi_1\approx \left[ 
      \begin{gathered} 
        0, \, \omega < \Omega; \\ 
        \pi,\, \omega > \Omega, \\ 
      \end{gathered} 
\right. \\[5mm]
\end{array}
\end{equation}
\begin{equation}\label{VEL21_0115}
 \sqrt{\omega^2-\eta^2}\approx \omega,\quad \sqrt{4\eta^2(\omega^2-\eta^2)+(\Omega^2-\omega^2+2\eta^2)^2}\approx \vert\Omega^2-\omega^2\vert.   
\end{equation}
Rewrite the wavenumber integral~(\ref{VEL21_1}), changing to the frequency integral:
\begin{equation}\label{VEL21_012}
\begin{array}{l}
     \DS v_{n}^\omega=\frac{2F_0\Omega e^{-\eta t} H(t)}{m\omega_e\pi}\int_0^{2\omega_e}\frac{(\Omega^2-\omega^2)\cos (\omega(\theta) t)}{(\Omega^2-\omega^2)^2+4\eta^2\Omega^2}\cos \left((2n+1)\arcsin\frac{\omega}{2\omega_e} \right) \mathrm{d}\omega.
     \end{array}{}
\end{equation}

Introduce a function, $\mathrm{I}_n^\mathrm{cr}$, which denotes the contribution to Eq.~(\ref{VEL21_1}), coming from the vicinity of the  point~$\Omega=\omega$. We introduce a variable~$\epsilon=\Omega-\omega$, which is infinitesimal for infinitesimal value $\eta/\omega_e$. Consider two cases:~$0<\Omega<2\omega_e$ and~$\Omega=2\omega_e$. In the first case, we transform~(\ref{VEL21_1}) using the following asymptotic expansions with respect to~$\epsilon/\omega_e$:
\begin{equation}\label{VEL21_2_1}
    \begin{array}{l}
     \DS \Omega^2-\omega^2=2\Omega\epsilon+O\left(\left(\frac{\epsilon}{\omega_e}\right)^2\right),\\[2mm]
    \end{array}
\end{equation}
\begin{equation}\label{VEL_21_2_2}
    \cos\left((2n+1)\arcsin \frac{\omega}{2\omega_e}\right)=\cos\left((2n+1)\arcsin \frac{\Omega}{2\omega_e}\right)+O\left(\frac{\epsilon}{\omega_e}\right).
\end{equation}
Based on aforesaid, we write equation for~$\mathrm{I}^\mathrm{cr}_n$ as the integral along the vicinity of~$\epsilon=0$:
\begin{equation}\label{VEL21_3}
   \begin{array}{l}
    \DS \mathrm{I}^\mathrm{cr}_n=\frac{F_0e^{-\eta t}H(t)\mathcal{T}_{2n+1}\left(\sqrt{1-\frac{\Omega^2}{4\omega_e^2}}\right)}{m\omega_e \pi}\Bigg[\cos(\Omega t)\int_{\Omega-2\omega_e}^{\Omega}\frac{\epsilon\cos(\epsilon t)\mathrm{d}\epsilon}{\epsilon^2+\eta^2}+\\[3mm]
    \DS \sin(\Omega t)\int_{\Omega-2\omega_e}^{\Omega}\frac{\epsilon\sin(\epsilon t)\mathrm{d}\epsilon}{\epsilon^2+\eta^2}\Bigg].
    \end{array}
\end{equation}
Introduce a function~$v^\omega_{n,\mathrm{cr}}$ such as
\begin{equation}\label{VEL21_4}
    v^\omega_{n,\mathrm{cr}}\approx\mathrm{I}^\mathrm{cr}_n \vert_{\Omega t\gg 1,\, \eta \rightarrow 0+}.
\end{equation}
Then, tending~$\epsilon t\rightarrow \infty$ in the integrals in Eq.~(\ref{VEL21_3}) and using the following identities, which are true for any~$\Tilde{\eta}>0$
\begin{equation}\label{VEL21_5}
    \int_{-\infty}^{\infty}\frac{x\sin x}{x^2+\Tilde{\eta}^2}\mathrm{d}x=e^{-\Tilde{\eta}}\pi,\qquad \int_{-\infty}^{\infty}\frac{x\cos x}{x^2+\Tilde{\eta}^2}\mathrm{d}x=0,
\end{equation}
 we write the expression for~$v^\omega_{n,\mathrm{cr}}$ in the form
\begin{equation}\label{VEL21_6}
    v^\omega_{n,\mathrm{cr}}=  \DS\frac{F_0\mathcal{T}_{2n+1}\left(\sqrt{1-\frac{\Omega^2}{4\omega_e^2}}\right)}{m\omega_e}\sin(\Omega t).
\end{equation}
Consider the case when~$\Omega=2\omega_e$. Putting~$\eta=0$ and ~$\epsilon=2\omega_e-\omega$, we rewrite the asymptotic expansions~(\ref{VEL_21_2_2}) as
\begin{equation}\label{VEL21_7_1}
     \DS \Omega^2-\omega^2=4\omega_e\epsilon+O\left(\left(\frac{\epsilon}{\omega_e}\right)^2\right),\\[2mm]
\end{equation}
\begin{equation}\label{VEL21_7_2}
    \DS \cos\left((2n+1)\arcsin \frac{\omega}{2\omega_e}\right)=(-1)^n(2n+1)\sqrt{\frac{\epsilon}{\omega_e}}+O\left(\left(\frac{\epsilon}{\omega_e}\right)^\frac{3}{2}\right).
\end{equation}
Therefore, 
\begin{equation}\label{VEL21_8}
\small
  \mathrm{I}^\mathrm{cr}_n=\frac{(-1)^nF_0H(t)(2n+1)}{m\pi\sqrt{\omega_e^3}}\Bigg[\cos(2\omega_e t)\int_{0}^{2\omega_e}\frac{\cos(\epsilon t)\mathrm{d}\epsilon}{\sqrt{\epsilon}}+\sin(2\omega_e t)\int_{0}^{2\omega_e}\frac{\sin(\epsilon t)\mathrm{d}\epsilon}{\sqrt{\epsilon}}\Bigg].
\end{equation}
\normalsize
Tending~$ \epsilon t\rightarrow \infty$ in~(\ref{VEL21_8}) and using Eq.~(\ref{VEL21_4}) and the identities
\begin{equation}\label{VEL21_9}
    \int_0^\infty \frac{\cos x\,\mathrm{d}x}{\sqrt{x}}=\sqrt{\frac{\pi}{2}},\qquad \int_0^\infty \frac{\sin x\mathrm{d}x}{\sqrt{x}}=\sqrt{\frac{\pi}{2}},
\end{equation}
we write the expression for~$v^\omega_{n,\mathrm{cr}}$ as 
\begin{equation}\label{VEL21_10}
    v^\omega_{n,\mathrm{cr}}\approx  \DS\frac{(-1)^nF_0(2n+1)}{m\sqrt{2\pi\omega_e^3 t}} \left(\cos(2\omega_e t)+\sin(2\omega_e t)\right),
\end{equation}
which can be transformed to the form shown in Eq.~(\ref{CritEq}). 

\section{Derivation of the expression~(\ref{StatEq})}\label{CCC}
Put~$\eta=0$ in~Eq.~(\ref{Vel2}) and rewrite Eq. for~$v_{n}^\omega$ in as follows:
\begin{equation}\label{VEL22_1}
   \DS v_{n}^\omega=\frac{F_0\Omega H(t)}{m}\left(\mathrm{I}_1-\mathrm{I}_2+\mathrm{I}_3+\mathrm{I}_4\right),
\end{equation}
\begin{equation}\label{VEL22_21}
    \mathrm{I}_{1,3}=\frac{1}{\pi}\int_0^{\pi}\frac{\cos(\omega(\theta)t\pm n\theta)}{\Omega^2-\omega(\theta)^2}\cos^2\frac{\theta}{2}\mathrm{d}\theta,\quad \mathrm{I}_{2,4}=\frac{1}{\pi}\int_0^{\pi}\frac{\sin(\omega(\theta)t\pm n\theta)}{\Omega^2-\omega(\theta)^2}\cos\frac{\theta}{2}\sin\frac{\theta}{2}\mathrm{d}\theta.
\end{equation}
In order to estimate the large-time asymptotics of the integrals in Eq.~(\ref{VEL22_1}) at the moving front, firstly, we transform them to the form with the structure of the Fourier integral:
\begin{equation}\label{VEL22_2}
    \Tilde{\mathrm{I}}=\int f(\theta)e^{\mathrm{i}\varphi(\theta) t}\mathrm{d}\theta.
\end{equation}
Following~\cite{Slepyan1972, whitham2011linear, Gavrilov2022}, put~$n=w\omega_e t$, where~$w$ is the dimensionless constant in time speed of the observation point. Then, we consider the following integrals:
\begin{equation}\label{VEL22_3}
    \DS \Tilde{\mathrm{I}}_{1\pm}=\frac{1}{\pi}\int_0^{\pi} f_{1}(\theta)e^{\mathrm{i}\varphi_\pm(\theta)\omega_e t}\mathrm{d}\theta,\qquad \Tilde{\mathrm{I}}_{2\pm}=\frac{1}{\pi}\int_0^{\pi} f_{2}(\theta)e^{\mathrm{i}\varphi_\pm(\theta)\omega_e t}\mathrm{d}\theta,
\end{equation}
\begin{equation}\label{VEL22_35}
     f_1(\theta)=\frac{\DS \cos^2\frac{\theta}{2}}{\Omega^2-\omega(\theta)^2},\qquad f_2(\theta)=\frac{\DS \cos\frac{\theta}{2}\sin\frac{\theta}{2}}{\Omega^2-\omega(\theta)^2},\qquad \varphi_\pm(\theta)=2\sin \frac{\theta}{2}\pm w\theta.
\end{equation}
For estimation of large-time asymptotics, we use the stationary phase method~\cite{Fedoryuk}. The stationary points, $\theta_s$, corresponding to the integrals in Eq.(\ref{VEL22_3}) satisfy the condition~$\DS \frac{\mathrm{d}\varphi_{\pm}}{\mathrm{d}\theta}\Big\vert_{\theta=\theta_s}=0$. Therefore, $0<w<1$. As well as the negative stationary points, the stationary point~$\theta_s=2(\pi-\arccos{w})$ does not belong to the interval~$(0; \pi)$. Therefore,
\begin{equation}\label{VEL22_4}
    \Tilde{\mathrm{I}}_{1+}=\Tilde{\mathrm{I}}_{2+}=O((\omega_et)^{-\infty})
\end{equation}
and thus~$\mathrm{I}_1=\mathrm{I}_2=O((\omega_et)^{-\infty})$. Consider~$\theta_s=2\arccos w$. Then
\begin{equation}\label{VEL22_5}
    \begin{array}{l}
     \DS \varphi_{-}(\theta_s)=2\sqrt{1-w^2}-2w\arccos w,\quad \DS \frac{\mathrm{d}^2\varphi_{-}}{\mathrm{d}\theta^2}\Big \vert_{\theta=\theta_s}=-\frac{1}{2}\sqrt{1-w^2}<0.
    \end{array}
\end{equation}
Since the stationary point is not degenerate, we can therefore write the following expression for the principal term of the asymptotics of the integral~$\Tilde{\mathrm{I}}_{1-}$~\cite{Fedoryuk}:
\begin{equation}\label{VEL22_6}
    \begin{array}{l}
    \DS \Tilde{\mathrm{I}}_{1-}\backsim\frac{1}{\pi}\sqrt{\frac{2\pi}{\omega_e t\Big \vert\frac{\mathrm{d}^2\varphi_{-}}{\mathrm{d}\theta^2}\big \vert_{\theta=\theta_s}\Big\vert}}f_1(\theta_s)e^{\mathrm{i}\left(\varphi_{-}(\theta_s)\omega_e t+\frac{\pi}{4}\mathrm{sgn} \frac{\mathrm{d}^2\varphi_{-}}{\mathrm{d}\theta^2}\Big \vert_{\theta=\theta_s}\right)}\\[3mm]
    \DS =\frac{2}{\sqrt{\pi \omega_e t\sqrt{1-w^2}}}\frac{w^2}{\Omega^2-4\omega_e^2(1-w^2)}e^{\mathrm{i}\left((2\sqrt{1-w^2}-2w\arccos w)\omega_et-\frac{\pi}{4}\right)}.
    \end{array}
\end{equation}
Analogously,
\begin{equation}\label{VEL22_7}
    \DS \Tilde{\mathrm{I}}_{2-}\backsim\frac{2}{\sqrt{\pi \omega_e t\sqrt{1-w^2}}}\frac{w\sqrt{1-w^2}}{\Omega^2-4\omega_e^2(1-w^2)}e^{\mathrm{i}\left((2\sqrt{1-w^2}-2w\arccos w)\omega_et-\frac{\pi}{4}\right)}.
\end{equation}
The integrals~$\mathrm{I}_3$ and~$\mathrm{I}_4$ are calculated as
\begin{equation}\label{VEL22_8}
\mathrm{I}_3=\Re(\Tilde{\mathrm{I}}_{1-})H(1-w),\qquad \mathrm{I}_4=\Im(\Tilde{\mathrm{I}}_{2-})H(1-w).
\end{equation}
Substitution of~$w=n/(\omega_et)$ to Eqs.~(\ref{VEL22_6}),~(\ref{VEL22_7}) and~(\ref{VEL22_8}) and the final result to Eq.~(\ref{VEL22_1}) with further simplifications yields the expressions~(\ref{StatEq}) \edited{and~(\ref{StatEq2})}. 

\section{Solution of Eq.~(\ref{EQ_MAX})}\label{DDD}
We seek solution of~Eq.~(\ref{EQ_MAX}) with accuracy up to order of~$\beta F_0^2/c^3$. Knowing that~$\Omega_\mathrm{cr}>2\omega_e$ but not much more, than~$2\omega_e$ we can write expression for~$\mu(\Omega_\mathrm{cr})$ as

\begin{equation}\label{ser_cr}
 \mu(\Omega_\mathrm{cr})=\mu(2\omega_e)+O \left(\frac{\beta F_0^2}{c^3}\right).
\end{equation}
Substituting~(\ref{ser_cr}) to~(\ref{EQ_MAX}) with preserving terms of order of~$\beta F_0^2/c^3$ yields
\begin{equation}\label{EQ_MAX_NEW}
    \Omega_\mathrm{cr}\left(1-\frac{\beta F_0^2}{c^3} \mu(2\omega_e)\right)=2\omega_e,
\end{equation}
whereas 
\begin{equation}\label{EQ_MAX_NEW2}
    \Omega_\mathrm{cr}=\frac{2\omega_e}{1-\frac{\beta F_0^2}{c^3} \mu(2\omega_e)}= 2\omega_e\left(1+\frac{\beta F_0^2}{c^3}\mu(2\omega_e)+O \left(\frac{\beta^2 F_0^4}{c^6}\right) \right).
\end{equation}
\end{appendices}


\begin{thebibliography}{9}

\bibitem{Pupin} Pupin, M. Propagation of long electrical waves. Transaction of the AIEE, pp.91-142~(1899)
\bibitem{mead1971} 
Mead, D.J. Vibration Response and Wave Propagation in Periodic Structures. Journal of Engineering for Industry, 93(3)~(1971) 
\bibitem{DNA} 
Svidlov, A., Drobotenko, M. et al. DNA dynamics under periodic force effects. International Journal of Molecular Sciences, 22(15)~(2021) 
\bibitem{Sinko} Shkurinov, A. P., Sinko, A. S. et al. Impact of the dipole contribution on the terahertz emission of air-based plasma induced by tightly focused femtosecond laser pulses. Physical Review E, 95(4), 043209~(2017)
\bibitem{Ovch} Ovchinnikov, A. A. Localized long-lived vibrational states in molecular crystals. Sov. Phys. JETP, 30(1), 147-150~(1970)
\bibitem{Sievers} Sievers, A., Takeno, S. Intrinsic Localized Modes in Anharmonic Crystals. Phys. Rev. Lett., 61(8)~(1988)
\bibitem{Flach} Flach, S., Willis, C. Discrete breathers. Physics Reports. Vol.295~(1998)
\bibitem{Kosevich} Kosevich, A. M., Kovalev, A. S. Selflocalization of vibrations in a one-dimensional anharmonic chain. Sov. Phys. JETP, 67, 1793~(1974)
\bibitem{Dolgov} Dolgov, A. S. On the localization of vibrations in a nonlinear crystal structure. Fizika Tverdogo Tela, 28(6), 1641-1644~(1986)
\bibitem{Sato} Sato, M., Mukaide, T. et al. Inductive intrinsic localized modes in a one-dimensional nonlinear electric transmission line. Phys.Rev.E, 94(1)~(2016)
\bibitem{PREBreathers} Saadatmand, D., Xiong, D., Kuzkin, V.A., Krivtsov, A.M., Savin, A.V., Dmitriev, S.V. Discrete breathers assist energy transfer to ac-driven nonlinear chains. Phys. Rev. E., 97(2)~(2018)
\bibitem{eva2017} Evazzade, I., Lobzenko, I., Korznikova, E. et al. Energy transfer in strained graphene assisted by discrete breathers excited by external ac driving. Phys. Rev. B., 95(3)~(2017)
\bibitem{Caputo} Caputo, J., Leon, J., Spire, A., et al. Nonlinear energy transmission in the gap. Phys. Rev. A., Vol.283, pp.129-135~(2001)
\bibitem{Watanabe2012} Watanabe, Y., Hamada, K., Sugimoto, N. Mobile intrinsic localized modes of a spatially periodic and articulated structure. Journal of the Physical Society of Japan, 81(1)~(2012)
\bibitem{Watanabe2015} Watanabe, Y., Nishida, T., Sugimoto, N. Excitation of intrinsic localized modes in finite mass-spring chains driven sinusoidally at end. Proceedings of the Estonian Academy of Sciences, 64(3)~(2015)
\bibitem{Watanabe2017} Watanabe, Y., Nishimoto, M., Shiogama, C. Experimental excitation and propagation of nonlinear localized oscillations in an air-levitation-type coupled oscillator array. Nonlinear Theory and Its Applications, IEICE, 8(2)~(2017)
\bibitem{Watanabe2018} Watanabe, Y., Nishida, T., Doi, Y., Sugimoto, N. Experimental demonstration of excitation and propagation of intrinsic localized modes in a mass–spring chain. Physics Letters, Section A: General, Atomic and Solid State Physics, 382(30)~(2018)
\bibitem{Geniet2002} Geniet, F., Leon, J. Energy transmission in the forbidden band gap of a nonlinear chain. Phys. Rev. Lett, 89(13)~(2002)
\bibitem{Geniet2003} Geniet, F., Leon, J. Nonlinear supratransmission. Journal of Physics: Condensed Matter, 15(17), 2933~(2003)
\bibitem{Leon} Leon, J. Nonlinear supratransmission as a fundamental instability. Physics Letters A, 319(1-2), 130-136~(2003)
\bibitem{Macias2008} Macías-Díaz, J. E. Numerical study of the transmission of energy in discrete arrays of sine-Gordon equations in two space dimensions. Physical Review E, 77(1), 016602~(2008)
\bibitem{Macias2016} Macías-Díaz, J. E. Numerical study of the process of nonlinear supratransmission in Riesz space-fractional sine-Gordon equations. Communications in Nonlinear Science and Numerical Simulation, 46, 89-102~(2017)
\bibitem{Santis} De Santis, D., Guarcello, C. et al. Supratransmission-induced travelling breathers in long Josephson junctions. Communications in Nonlinear Science and Numerical Simulation, 115, 106736~(2022)
\bibitem{Santis1} De Santis, D., Guarcello, C. et al. Generation of travelling sine-Gordon breathers in noisy long Josephson junctions. Chaos, Solitons and Fractals, 158, 112039~(2022)
\bibitem{Santis2} De Santis, D., Guarcello, C. et al. Breather dynamics in a stochastic sine-Gordon equation: evidence of noise-enhanced stability. Chaos, Solitons and Fractals, 168, 113115~(2023)
\bibitem{Susanto} Susanto, H. Boundary driven waveguide arrays: Supratransmission and saddle-node bifurcation.  arXiv:0809.3861~(2008)
\bibitem{Motcheyo2} Motcheyo, A., Kimura, M. Doi, Y et al.
Supratransmission in discrete one-dimensional lattices with the cubic–quintic nonlinearity. Nonlinear Dynamics, 95(3), pp.2461--2468~(2019)
\bibitem{Khomeriki} Khomeriki, R., Lepri, S., Ruffo, S. Nonlinear supratransmission and bistability in the Fermi-Pasta-Ulam model. Phys. Rev. E, 70(6)~(2004)
\bibitem{Khomeriki2} Khomeriki, R., Leon, J., Chevriaux, D.: Quantum hall bilayer digital amplifier. The European Physical Journal B-Condensed Matter and
Complex Systems 49, 213–218~(2006)
\bibitem{Kenmogne} Kenmogne, F. et al. Nonlinear supratransmission in a discrete nonlinear electrical transmission line: Modulated gap peak solitons. Chaos, Solitons and Fractals. Vol. 75, pp.263-271~(2015)
\bibitem{Motcheyo} Motcheyo, A., Tchawoua, C., Tchameu, J. Supratransmission induced by waves collisions in a discrete electrical lattice. Phys. Rev. E., 88(4)~(2013)
\bibitem{Macias2018} Macías-Díaz, J.E., Bountis, A. Supratransmission in $\beta$-Fermi–Pasta–Ulam chains with different ranges of interactions. Communications in Nonlinear Science and Numerical Simulation, Vol.63, pp.307--321~(2018)
\bibitem{Macias2020} Macías-Díaz, J.E. Modified Hamiltonian Fermi–Pasta–Ulam–Tsingou arrays which exhibit nonlinear supratransmission. Results in Physics, Vol.18, pp.1--11~(2020)
\bibitem{Gendelman2022} Bader, A., Gendelman, O. V. Supratransmission in a vibro-impact chain. Journal of Sound and Vibration, 547, 117493~(2023)
\bibitem{Kuz2018} Kuzkin, V.A., Krivtsov, A.M. Energy transfer to a harmonic chain under kinematic and force loadings: Exact and asymptotic solutions. Journal of Micromechanics and Molecular Physics, 3(1-2)~(2018)
\bibitem{Cannas1991} Cannas, S., Prato, D. Externally excited semi‐infinite one‐dimensional models. American Journal of Physics, 59(10), pp.915--920~(1991)
\bibitem{mokole1990exact} Mokole, E. L., Mullikin, A. L., Sledd, M. B. Exact and steady‐state solutions to sinusoidally excited, half‐infinite chains of harmonic oscillators with one isotopic defect. Journal of mathematical physics, 31(8), 1902-1913~(1990)
\bibitem{cherednichenko2019nonlinear} Cherednichenko, A.I., Zakharov, P.V., Starostenkov, M.D., Sysoeva,
M.O., Eremin, A.M.: Nonlinear supratransmission in a pt 3 al crystal at intense external influence~(in russian). Computer research and modeling
11(1), 109–117~(2019)
\bibitem{Dhar} Dhar, A. Heat transport in low-dimensional systems. Advances in Physics, 57(5)~(2008)
\bibitem{ChenG} Chen, G. Non-Fourier phonon heat conduction at the microscale and nanoscale. Nature Reviews Physics, 3(8)~(2021)
\bibitem{Lepri2016} Lepri, S., Livi, R., Politi, A. (2016). Heat Transport in Low Dimensions: Introduction and Phenomenology. In: Lepri, S. (eds) Thermal Transport in Low Dimensions. Lecture Notes in Physics, vol 921. Springer, Cham.
\bibitem{Podolskaya} Podolskaya, E. A., Krivtsov, A. M. and Kuzkin, V. A. Discrete thermomechanics: From thermal echo to ballistic resonance (a review). Mechanics and Control of Solids and Structures, 501-533~(2022)
\bibitem{Lepri2023} Benenti, G., Donadio, D., Lepri, S. and Livi, R. Non-Fourier heat transport in nanosystems. La Rivista del Nuovo Cimento, 1-57~(2023)
\bibitem{ChenJ} Bao, H., Chen, J., Gu, X., Cao, B. A review of simulation methods in micro/nanoscale heat conduction. ES Energy Environment, 1(39), 16-55~(2018)
\bibitem{ai2010heat} Ai, B.-q., He, D., Hu, B., et al.: Heat conduction in driven frenkel-
kontorova lattices: Thermal pumping and resonance. Physical Review E
81(3), 031124~(2010)
\bibitem{ZAMM} Krivtsov, A. M. Dynamics of matter and energy. ZAMM‐Journal of Applied Mathematics and Mechanics/Zeitschrift für Angewandte Mathematik und Mechanik, e202100496~(2022)
\bibitem{Vladimirov} Vladimirov, V.: Equations of Mathematical Physics. Marcel Dekker, New York~(1971)
\bibitem{Indeitsev1999} Alekseev, V. V., Indeitsev, D. A., Mochalova, Y. A. Resonant oscillations of an elastic membrane on the bottom of a tank containing a heavy liquid. Technical Physics, 44, 903-907~(1999)
\bibitem{Lee} Lee, K. Propagation of a general disturbance along a semi-infinite linear chain. American Journal of Physics, 40(7), pp.1032--1034~(1972)
\bibitem{Lee1972} Lee, K., Kim, H. Exact Solutions for Dynamics of Finite, Semi-Infinite, and Infinite Chains with General Boundary and Initial Conditions. J. Chem. Phys. 57, 5037~(1972)
\bibitem{Nay} Nayfeh, A., Rice, M. On the propagation of disturbances in a semi-infinite one-dimensional lattice. American Journal of Physics, 40(3), pp.469--470~(1972)
\bibitem{Prato} Prato, D., Lamberti, P. Quantum dynamics of a semi-infinite homogeneous harmonic chain. Physica A, 197, pp.232--242~(1993)
\bibitem{Ahmed} Ahmed, H., Nataryan, T., Rao, K.R. Discrete cosine transform. IEEE Transactions on Computers C-23(1), 90–93~(1974)
\bibitem{Hemmer} Hemmer, P. C. Dynamic and stochastic types of motion in the linear chain. Tapir forlag~(1959)
\bibitem{Gavr2023} Shishkina, E. V., Gavrilov, S. N. Unsteady ballistic heat transport in a 1D harmonic crystal due to a source on an isotopic defect. Continuum Mechanics and Thermodynamics, 35(2), 431-456~(2023)
\bibitem{shishkina2023localized} Shishkina, E.V., Gavrilov, S.N.: Localized modes in a 1d harmonic crys-
tal with a mass-spring inclusion. In: Advances in Linear and Nonlinear
Continuum and Structural Mechanics, pp. 461–479. Springer, Cham~(2023)
\bibitem{Gavr1999} Gavrilov, S. N. Non-stationary problems in dynamics of a string on an elastic foundation subjected to a moving load. J. Sound Vib. 222(3), 345–361~(1999)
\bibitem{SlepyanTsareva} Slepyan, L. I., Tsareva, O. V. Energy flux for zero group velocity of the carrying wave. In Soviet Physics Doklady. Vol. 32, p. 522~(1987)
\bibitem{Narisetti} Narisetti, R., Leamy, M., Ruzzene, M. A perturbation approach for predicting wave propagation in one-dimensional nonlinear periodic structures. Journal of Vibration and Acoustics, Transactions of the ASME. 132(3)~(2010)
\bibitem{Sepehri} Sepehri, S., Mashhadi, M. M. and Fakhrabadi, M. M. S. Dispersion curves of electromagnetically actuated nonlinear monoatomic and mass-in-mass lattice chains. International Journal of Mechanical Sciences, 214, 106896~(2022)
\bibitem{Kolm1992} Zakharov, V., Lvov, V., Falkovich, G. Kolmogorov spectra of turbulence I. Wave turbulence.  Springer Series in Nonlinear Dynamics~(1992)
\bibitem{Shirokov} Shirokoff, D. Renormalized waves and thermalization of the Klein-Gordon equation. Physical Review E, 83(4), 046217~(2011)
\bibitem{Gershgorin2005} Gershgorin, B., Lvov, Y. V., Cai, D. Renormalized waves and discrete breathers in $\beta$-Fermi-Pasta-Ulam chains. Phys. Rev. Lett, 95(26), 264302~(2005)
\bibitem{Gershgorin2007} Gershgorin, B., Lvov, Y. V. and Cai, D. Interactions of renormalized waves in thermalized Fermi-Pasta-Ulam chains. Physical Review E, 75(4), 046603~(2007)
\bibitem{LeeW} Lee, W., Kovačič, G. and Cai, D. Renormalized resonance quartets in dispersive wave turbulence. Physical review letters, 103(2), 024502~(2009)
\bibitem{Kovacic} Brennan, M. J., Kovacic, I. The Duffing Equation: Nonlinear Oscillators and Their Phenomena. Wiley~(2011)
\bibitem{Panovko} Panovko, Y. G. A review of applications of the method of direct linearization. Applied Mechanics, pp.167--171~(1966)
\bibitem{HB} Hu, H. Solution of a quadratic nonlinear oscillator by the method of harmonic balance. Journal of Sound and Vibration, 293(1-2), pp.462--468~(2006)
\bibitem{KeenB} Keen, B., Fletcher, W. H. Nonlinear plasma instability effects for subharmonic and harmonic forcing oscillations. Journal of Physics A: General Physics, 5(1), 152~(1972)
\bibitem{Kuz2023} Kuzkin, V.A. Acoustic transparency of the chain-chain interface. Phys. Rev. E 107, 065004~(2023)
\bibitem{Terraneo} Terraneo, M., Peyrard, M., Casati, G. Controlling the energy flow in nonlinear lattices: a model for a thermal rectifier. Physical Review Letters, 88(9), 094302~(2002)
\bibitem{Kobayashi} Kobayashi, W., Teraoka, Y., Terasaki, I. An oxide thermal rectifier. Applied Physics Letters, 95(17), 171905~(2009)
\bibitem{Li2012} Li, N., Ren, J., Wang, L., Zhang, G., Hänggi, P., Li, B. Colloquium: Phononics: Manipulating heat flow with electronic analogs and beyond. Reviews of Modern Physics, 84(3), 1045~(2012)
\bibitem{Malik} Malik, F. K., Fobelets, K. A review of thermal rectification in solid-state devices. Journal of Semiconductors, 43(10), 103101~(2022)
\bibitem{GelfandShilov} Gelfand, I., Shilov, G.: Generalized Functions. Properties and Operations,
vol. 1. Academic Press, New York~(1964)
\bibitem{Slepyan1972} Slepyan, L. I. Nonstationary elastic waves. Sudostroenie, Leningrad, 376~(1972) (in Russian)
\bibitem{whitham2011linear} Whitham, G.B.: Linear and Nonlinear Waves. John Wiley and Sons, New
York~(1974)
\bibitem{Gavrilov2022} Gavrilov, S.N. Discrete and continuum fundamental solutions describing heat conduction in
a 1D harmonic crystal: Discrete-to-continuum limit and slow-and-fast motions decoupling. International Journal of Heat and Mass Transfer, 194, 123019~(2022)
\bibitem{Fedoryuk} Fedoryuk, M. V. The saddle-point method. Science~(1977) (in Russian)





\end{thebibliography}
\end{document}